\documentclass[11pt]{article}
\usepackage{cite}
\usepackage{amsmath,amsfonts,amssymb}
\usepackage{slashed}

\def\hybrid{
        \topmargin -20pt
        \oddsidemargin 0pt
        \headheight 0pt \headsep 0pt
        \textwidth 6.25in % A4 paper
        \textheight 9.5in % A4 paper
        \marginparwidth .875in
        \parskip 5pt plus 1pt \jot = 1.5ex}

\hybrid

 \csname
@addtoreset\endcsname{equation}{section}

\def\moth{\mathsurround=0pt}
\newdimen\zo \zo=0pt

\def\tick{\leaders\hrule height 0.5ex depth 0pt \hskip 0.5pt}
\def\upboxfill{$\moth \setbox\zo\hbox{\tick}%
  \hskip 3pt\hbox to 0pt{$\tick$\hss}\hrulefill \hbox to 7.5pt{$\tick$\hss}$}

\def\dtick{\leaders\hrule height .34pt depth 0.5ex \hskip 0.5pt}
\def\downboxfill{$\moth \setbox\zo\hbox{\dtick}%
  \hskip 2pt\hbox to 0pt{$\dtick$\hss}\hrulefill \hbox to 2pt{$\dtick$\hss}$}

\def\ciftS{\mathbb{S}}
\def\cL{{\cal L}}
\def\cH{{\cal H}}
\def\cK{{\cal K}}
\def\cR{{\cal R}}

\def\hl{\hat\lambda}

\def\hl{\hat\lambda}

\def\del{\partial}

\def\be{\begin{equation}}
\def\ee{\end{equation}}
\def\bea{\begin{eqnarray}}
\def\eea{\end{eqnarray}}
\def\ba{\begin{array}}
\def\ea{\end{array}}

\begin{document}

\begin{titlepage}
\rightline{} \rightline\today
\begin{center}
\vskip 2.5cm {\Large \bf { Massive Deformations of Type IIA  Theory Within Double Field Theory}}\\
\vskip 2.5cm {\large {Aybike \c{C}atal-\"{O}zer}} \vskip 1cm
{\it {Department of Mathematics,}}\\
{\it {\.{I}stanbul Technical University,}}\\
{\it {Maslak 34469,
\.{I}stanbul, Turkey}}\\
ozerayb@itu.edu.tr \vskip 0.5cm

\vskip 1cm {\bf Abstract}
\end{center}

\vskip 0.5cm

\noindent
\begin{narrower}

\noindent We  obtain  massive deformations of Type IIA
supergravity theory through  duality twisted reductions of Double
Field Theory (DFT) of massless Type II strings. The mass
deformation is induced through the reduction of the DFT of the RR
sector. Such reductions are determined by a twist element
belonging to $Spin^+(10,10)$, which is the duality group of the
DFT of the RR sector.  We determine the form of the twists and
give particular examples of twist matrices, for which a massive
deformation of Type IIA theory can be obtained. In one of the
cases, requirement of gauge invariance of the RR sector implies
that the dilaton field must pick up a linear dependence on one of
the dual coordinates. In another case, the choice of the twist
matrix violates the weak and the strong constraints explicitly in
the internal doubled space.

\end{narrower}

\vspace{4cm}

\end{titlepage}

\newpage

\tableofcontents

\section{Introduction}

Double Field Theory (DFT) is a field theory defined on a doubled
space, which implements the $O(d,d)$ T-duality symmetry of string
theory as a  manifest symmetry. In addition to the standard
space-time coordinates, the doubled space also includes dual
coordinates, which are associated with the  winding excitations of
closed string theory on backgrounds with non-trivial cycles. The
DFT action was constructed in \cite{HullZ1,HullZ2,HullZ3,HullZ4},
building on earlier work
\cite{Tseytlin1,Tseytlin2,Siegel1,Siegel2,Siegel3,Hulleski1,Hulleski2,Hulleski3,Hulleski4}.
The consistency of the action requires the imposition of a set of
constraints, called the weak and the strong constraints. The
strong constraint implies that locally all fields and parameters
of the theory depend only on half of the coordinates. In a certain
frame called the supergravity frame, in which none of the fields
depend on the dual coordinates associated with the winding
excitations, the DFT action constructed by
\cite{HullZ1,HullZ2,HullZ3,HullZ4} reduces to the action of the
NS-NS sector of massless string theory. The weak and the strong
constraints are satisfied in a trivial way for this choice of
frame. The gauge transformation rules of DFT are governed by the
so-called C-bracket, which is an $O(d,d)$ covariantization of the
Courant bracket in generalized geometry of Hitchin
\cite{Hitchin,Gualtieri}. In later work \cite{dftRR,dftRRkisa},
the DFT action of the RR sector of Type II string theory was also
constructed\footnote{An alternative formulation of the RR sector,
called the semi-covariant formulation is given in the papers
\cite{park1,park2}.}, where it was  shown that in the supergravity
frame this new action reduces to the democratic formulation of the
RR sector of Type II supergravity. The fields in the RR sector are
chiral spinor fields, which form a representation of $Pin(10,10)$.
The action does not possess the full $Pin(10,10)$ symmetry. The
chirality condition and the existence of a self-duality condition
which must supplement the action reduces the symmetry group to
$Spin^+(10,10)$.

In the interesting paper \cite{HohmKwak}, the action constructed
by \cite{dftRR} was shown to give rise to massive Type IIA theory
of \cite{Romans} in a certain frame, in which the RR one-form of
Type IIA is allowed to pick up a linear dependence on one of the
dual coordinates. Unlike the supergravity frame, in which the
action reduces to the democratic formulation of massless type II
theories, the strong constraint is violated in the frame
considered in \cite{HohmKwak}. However, the weak constraint is
still respected. To be more precise, the anzats (through which the
RR fields pick up their coordinate dependencies) of
\cite{HohmKwak} is not in the most general possible form allowed
by the weak constraint, and owes its consistency  to the fact that
the dependence on the dual coordinates is  linear. Consistency of
the theory also requires a reformulation of the gauge
transformations on the RR fields, in such a way that the linear
dependence on the dual coordinates would drop out. The work of
\cite{HohmKwak} is particularly interesting, as it also addresses
the challenge of relaxing the strong constraint.

There has been other attempts  to relax the strong and the weak
constraints by various groups. One line of work in this direction
involves formulating the DFT action on group manifolds rather than
toroidal backgrounds, as initiated by \cite{hassler1,hassler2}.
Possibility of relaxing the strong constraint, albeit  partially,
also comes from  the study of duality twisted reductions of DFT.
As mentioned above, DFT comes equipped with the large duality
group $O(d,d)$, which makes it possible to introduce a duality
twisted reduction anzats (also called generalized Scherk-Schwarz
type anzats \cite{SS}) for the fields and gauge parameters of the
theory. This line of work has been pursued by many groups so far
\cite{Grana,Geissbuhler,Aldazabal,parktwist,berman,Lust,Geissbuhler2}.
An interesting aspect of these works is that one never needs to
impose the weak and the strong constraints
 on the doubled internal space.
This feature was made  explicit  in \cite{Grana}, where they gave
the set of conditions to be satisfied for the consistency of the
twisted reduction and showed that these constituted a weaker set
of constraints compared with the  constraints of DFT.

The linear anzats of \cite{HohmKwak} and the Scherk-Schwarz type
anzatse mentioned above are two prominent examples with success
towards the aim of relaxing the constraints of DFT. This has
motivated us to study these two approaches together. In this
paper, our aim is to explore whether massive deformations of Type
IIA theory can be obtained through duality twisted reductions of
DFT. Indeed, we show that a mass deformation can be induced
through the reduction of the DFT of the RR sector. Such reductions
have been studied recently \cite{Aybike} and are determined by a
twist element belonging to $Spin^+(10,10)$, which is the duality
group of the DFT of the RR sector. We determine  the form of the
twists and give particular examples of twists matrices, for which
a  massive deformation of Type IIA theory can be obtained. In one
of the cases, requirement of gauge invariance of the RR sector
implies that the dilaton field must pick up a linear dependence on
one of the dual coordinates. We find this aspect particularly
interesting, given that reductions with non-trivial dilaton anzats
have received some interest recently, in  attempts to understand
the recently discovered generalized supergravity equations
\cite{Tseytlin} within DFT \cite{Yoshida2,Yoshida1,Samtleben}. In
another case, the choice of the twist matrix violates the weak and
the strong constraints explicitly in the internal doubled space.

The plan of the paper is as follows. In the next section, we give
a brief review of the Double Field Theory of  Type II strings and
its reduction with duality twists. Section (\ref{sectionmassive})
is the main section of the paper, where we determine the twist
element, which gives rise to the massive deformation of  Type IIA
theory. In order to explain how the mass deformation arises, we
start by a more detailed explanation of the DFT action of the  RR
sector, focusing on its formulation in terms of Mukai pairing,
which is a $Spin(d,d)$ invariant bilinear form on the space of
spinors \cite{Mukai}. Then, we continue with the computation of
fluxes that determine the deformations of the RR sector. This
makes it possible to determine the possible forms of twist that
could give rise to a mass deformation in the subsequent
subsections. We end the paper with discussions and outlook in
section (\ref{conclusion}). In the main body of the paper, we
mainly focus on the RR sector, as the mass deformation is induced
through the reduction of this sector. The Appendix
(\ref{Appendix}) collects useful information about the NS-NS
sector and its deformation through duality twisted reduction
anzatse.

%%%%%%%%%%%%%%%%%%%%%%%%%%%%%%%%%%%%%%%%%%%%%%%%%%%%%%%%

\section{Review of Duality Twisted Reductions of
DFT}\label{review}

In this section, we give a very brief review of DFT and its
duality twisted reduction. Our aim here is to set the notation and
present the formulas needed in the main section
(\ref{sectionmassive}). For a more complete introduction to these
topics, see
\cite{HullZ3,dftRR,Aldazabal,Grana,Geissbuhler2,Aybike}.

In DFT,   the number of  space-time coordinates is doubled by
introducing  dual coordinates  $\tilde{x}_i$, which are associated
with winding excitations of string theory on backgrounds with
non-trivial cycles. The standard and the dual coordinates combine
to form an $O(d,d)$ vector transforming as: \be X'^M =
h^M_{\ \ N} X^N, \ \ \ \ X^M =  \left(\ba{c} \tilde{x}_i \\
                                              x^i \ea \right) \ee
Here  $h^M_{\ \ N}$ is a general $O(d,d)$ matrix.  We  raise and
lower indices by the $O(d,d)$ invariant metric $\eta$, and hence
$X_M = \eta_{MN} X^N$. The $O(d,d)$ invariant constraints are
  \be\label{ODDconstr}
   \partial^{M}\partial_{M}A \ = \ \eta^{MN}\partial_{M}\partial_{N}A \ = \ 0\;, \qquad
   \partial^{M}A\,\partial_{M}B \ = \ 0\;, \qquad
   \eta^{MN} \ = \  \begin{pmatrix}
    0&1 \\1&0 \end{pmatrix}\,,
 \ee
where $A$ and $B$ represent any fields or parameters of the
theory. The first of the above constraints is called the
\emph{weak constraint}. It follows from the level matching
constraint in closed string theory. The second constraint is
stronger and is called the \emph{strong constraint}. It is too
strong actually, as it implies that all fields and gauge
parameters of the theory can depend only on half of the
coordinates.

The DFT action of Type II strings is \cite{HullZ4,dftRR,dftRRkisa}
\be \label{DFTaction} {\cal{S}} = \int \ dx d\tilde{x} \left(
{\cal{L}}_{{\rm NS-NS}} + {\cal{L}}_{{\rm RR}} \right), \ee where
\be \label{DFTaction1} {\cal{L}}_{{\rm NS-NS}} = e^{-2 d} \
{\cal{R}} (\cH, d) \hspace*{1cm} {\rm and} \hspace*{1cm}
{\cal{L}}_{{\rm RR}} =  \frac{1}{4} (\slashed{\partial}
\chi)^{\dagger} \ {\mathbb{S}} \ \slashed{\partial} \chi =
\frac{1}{4} \langle  \ \slashed{\partial} \chi, \ C^{-1} \ciftS \
\slashed{\partial} \chi \rangle.
\end{equation} Here, $\langle ~ \rangle$ is the Mukai pairing,
which is a $Spin(d,d)$ invariant bilinear form on the space of
spinors \cite{Mukai,Gualtieri}.  The  action (\ref{DFTaction}) has
to be supplemented by the following self-duality constraint \be
\label{selfduality} \slashed{\partial} \chi = - \cK \
\slashed{\partial}\chi, \ \ \ \cK \equiv C^{-1} \ciftS. \ee In the
supergravity frame in which $\tilde{\partial}^i = 0$,  the first
term in the above action reduces to the standard NS-NS action for
the massless fields of string theory and the second term reduces
to the RR sector of the democratic formulation of Type II
supergravity theories. For this reason, the first term is called
the DFT action of the NS-NS sector of string theory, whereas the
second term is
 referred to as the DFT action of the RR sector.

The $\cR(\cH, d)$ in (\ref{DFTaction1}) is the generalized Ricci
scalar \cite{HullZ4}. Its explicit form can be found in
(\ref{ricci}). The generalized Ricci scalar is defined in terms of
the generalized metric $\cH$ and the generalized dilaton field
$d$. We present these generalized $O(d,d)$ tensors  below.
\begin{equation}\label{genmetric}
 \cH_{MN} = \left(\begin{array}{cc}
                         \cH^{ij} & \cH^i_{\  j} \\
                         \cH_i^{\  j} & \cH_{ij} \\
                         \end{array}\right) = \left(\begin{array}{cc}
                         g^{ij}  & -g^{ik} b_{kj} \\
                         b_{ik}g^{kj} & g_{ij} - b_{ik} g^{kl} b_{lj}\\
                         \end{array}\right), \ \ \ \ e^{-2 d} =
\sqrt{g} e^{-2 \phi},
\end{equation} where, $\ g = \mid {\rm det} g \mid $. The generalized metric $\cH$ transforms covariantly under $O(d,d)$
transformations and the generalized dilaton field is invariant
under such transformations.

The  field $\ciftS$ in the  Ramond-Ramond sector is the spinor
representative of the generalized metric $\cH$ under the double
covering homomorphism $\rho: Spin(d,d) \rightarrow SO(d,d)$, that
is, $\rho(\ciftS) = \cH$. Since the generalized metric $\cH$ is in
the coset $SO^-(d,d)$ in Lorentzian signature, the spinor field
$\ciftS$ it lifts to, is an element of $Spin^-(d,d)$. The field
$\chi$, which is the other dynamical field in the DFT of the RR
sector is  a chiral spinor field. It encodes all the (modified)
p-form fields in the RR sector. The chirality of $\chi$ determines
whether the corresponding theory is Type IIA or Type IIB string
theory. The former corresponds to negative chirality of $\chi$,
whereas the latter involves the spinor field $\chi$ with postive
chirality. The operator $\slashed{\partial}$ in the action
(\ref{DFTaction}) is the generalized Dirac operator. It is defined
as \be \label{diracop} \slashed{\partial} \equiv
\frac{1}{\sqrt{2}} \Gamma^M
\partial_M = \frac{1}{\sqrt{2}}(\Gamma^i \partial_i + \Gamma_i
\tilde{\partial}^i) \equiv \psi^i \partial_i + \psi_i
\tilde{\partial}^i. \ee Here the Gamma matrices $\Gamma^M =
(\Gamma_i, \Gamma^i)$ are the matrix representations of the
generators of the Clifford algebra $CL(R^{2d}, \eta)$: \be
\label{clifford} \{\Gamma^M, \Gamma^N \} = 2 \eta^{MN}. \ee The
labelling is such that the first $d$ elements span a maximally
isotropic subsace with respect to the metric $\eta$ and the the
remaining elements span the orthogonal complement. Then one has
\be \label{fermosc}
  \{ \Gamma_i ,  \Gamma^j\} \ = \ 2 \delta_{i}{}^{j}\;, \qquad
  \{\Gamma_i,\Gamma_j\} \ = \  0\;, \qquad \{\Gamma^i,\Gamma^j\} \ = \ 0\;,
 \ee

 The algebra elements $\Gamma^M$ do not lie in the associated group
$Spin(d,d)$, so it is more appropriate to work with $\psi^M =
\frac{1}{\sqrt{2}} \Gamma^M$, which do lie in the spin
group\footnote{$\psi^M$ satisfy $\psi^M (\psi^M)^* = \pm 1$, so
they lie in the spinor group \cite{Gualtieri,Aybike}.}. The
spinorial action of the Clifford algebra elements $\Gamma^M$ on
spinor fields $\chi$ is best understood when one identifies the
spinor fields with (not necessarily homogenous) differential forms
(or with polyforms, as is commonly called in physics literature)
of the exterior algebra $\bigwedge^\bullet R^d$
\cite{Gualtieri,dftRR,Aybike}.\footnote{Note that, here  we work
 locally and work with the Clifford algebra on $R^{2d}$. A
more careful analysis would require to work on the $2d$
dimensional tangent space of the doubled manifold and discuss how
the local structure can be transported to the whole doubled
manifold. For a more detailed discussion, see \cite{Aybike}.}
Then, if one writes the exterior algebra element $\chi$ as \be
\label{spinorform} \chi(x, \tilde{x}) = \sum_p \frac{1}{p!} C_{i_1
\ldots i_p} (x, \tilde{x}) \psi^{i_1} \ldots \psi^{i_p}, \ee then
$\psi^i$ acts on $\chi$ by wedge product, whereas $\psi_i$ acts by
contraction. More, precisely, one has \be \label{gammaact} \psi^i
. \chi = \psi^i \wedge \chi, \ \ \ \psi_i . \chi = i_{\psi_i}
\chi, \ee where one defines $i_{\psi_i} \psi^j =
\delta_{i}{}^{j}$. For more details see \cite{dftRR,Aybike}.

The gauge transformation rules of the DFT action are determined by
the  generalized Lie derivative $\hat{\cL}$.  Let us define  the
$O(d,d)$ vector $\xi^M = (\tilde{\xi}_i,\xi^i)$ as the parameter
of the gauge transformations. Then, we have \cite{HullZ2} \bea
\label{gaugechi} ~
 \delta_\xi \chi \ = \ \widehat{\cal L}_\xi \chi  \ & \equiv & \ \xi^M \partial_M  \chi
\ + \   {1\over \sqrt{2}}\, \slashed{\partial} \xi^M
\Gamma_{\hskip-1pt M}\, \chi  \, \nonumber \eea \be \label{gaugeK}
\delta_{\xi} {\cal K}  \ = \ \xi^M
\partial_M  {\cal K} + {1 \over 2} \big[ \Gamma^{PQ}  , \,  {\cal
K} \, \big]  \partial_{P} \xi_{Q} \; , \ee in the RR sector, where
$\Gamma^{PQ} \equiv \frac{1}{2}  [\Gamma^P , \Gamma^Q ]$. The
double field theory version of the abelian gauge symmetry of
p-form gauge fields is \be \label{gaugelambda} \delta_{\lambda}
\chi = \slashed\del \lambda = \frac{1}{\sqrt{2}} \Gamma^M
\partial_M \lambda.\ee Here, $\lambda $ is a space-time dependent
spinor. The gauge transformation rules for the NS-NS sector is
given in
 Appendix \ref{Appendix}. The gauge algebra closes with respect to the
 C-bracket: \be
\label{cbracket} \big[ \xi_1, \ \xi_2 \big]_C^M = 2 \xi_{[ 1}^N
\del_N \xi_{2 ]}^M - \xi_{[ 1}^P \del^M \xi_{2 ]  P}. \ee The
C-bracket is the $O(d,d)$ covariantization of the Courant bracket
in generalized geometry \cite{Hitchin,Gualtieri}.

 The DFT action presented in (\ref{DFTaction}) is
invariant under the following transformations: \be
\label{transform}
 \ciftS(X) ~~  \longrightarrow \ciftS^{\prime}(X^{\prime}) \ = \ (S^{-1})^{\dagger}\, \ciftS(X)
 \,S^{-1}\;, ~~ \chi(X) \longrightarrow \chi(X') = S \chi(X)
 \ee
Here $S \in Spin^+(d,d)$  and $X^{\prime}= hX$,  where $h =
\rho(S) \in SO^+(d,d)$. The dilaton field is invariant. Although
the DFT action would be invariant under the above transformations
with a general $Pin(d,d)$ element $S$, the chirality of the spinor
field $\chi$ is preserved only by the $Spin(d,d)$ subgroup. On the
other hand, in order to preserve the self-duality constraint
(\ref{selfduality}), the duality group should be further reduced
to  $Spin^+(d,d)$. This is the reason why the actual duality group
in (\ref{transform}) is
 $Spin^+(d,d)$. The gauge transformation rules for the
generalized metric $\cH = \rho(\ciftS)$ is determined by those of
$\ciftS$ and is as given below: \be \label{transform2}
 \cH(X) ~~  \longrightarrow \cH^{\prime}(X^{\prime}) \ = \ (h^{-1})^{T}\, \cH(X)
 \,h^{-1}\;.
 \ee
 In determining the reduction
anzats, we also take into account the  global shift symmetry $d
\rightarrow d + \rho$ of the dilaton field, which acts as a
conformal rescaling on the NS-NS sector \cite{Grana}. This
symmetry extends to the RR sector too, provided that the spinor
field $\chi(X)$ transforms as $\chi \rightarrow e^{-\rho} \chi$
\cite{Aybike}. Finally, the DFT Lagrangian (\ref{DFTaction1}) is
also invariant under the  shift of the spinor field $\chi
\rightarrow \chi + \alpha$, where $\alpha$ is a constant spinor
field. Although this symmetry is not respected by the gauge
sector, it is still possible to take it into account in the
reduction anzats, as was discussed in \cite{Aybike}. All
symmetries considered, one ends up with the following duality
twisted anzats \cite{Aybike}: \bea \label{anzats}
\ciftS(X, Y) &=& (S^{-1})^{\dag}(Y) \ciftS(X) S^{-1}(Y), \\
\label{anzatschi} \chi(X, Y) &=& e^{-\rho(Y)} S(Y) (\chi(X) + \alpha), \\
 \label{anzatsdilaton} d(X, Y) &=& d(X) + \rho(Y). \eea
Here, $X$ denote collectively the coordinates of the reduced
theory. The $Y$  coordinates are the internal coordinates, which
will be integrated out eventually. One can further decompose these
coordinates into dual and standard coordinates as $Y = (\tilde{y},
y)$ and $X = (\tilde{x}, x)$.  The twist matrix $S(Y) \in
Spin^+(d,d)$ encodes the whole dependence of the fields and the
gauge parameters on the internal coordinates. Obviously, the above
anzats implies the following anzats in the NS-NS sector: \be
\label{anzats2} \cH_{MN}(X, Y) = U^A_{\ M}(Y) \cH_{AB}(X) U^B_{\
N}(Y).
 \ee

 The duality twisted dimensional reduction of the DFT action of the
NS-NS sector with the anzats
(\ref{anzats},\ref{anzatschi},\ref{anzatsdilaton}) and
(\ref{anzats2})
 has
been studied by several groups \cite{Geissbuhler,Aldazabal,Grana}.
The resulting theory was dubbed Gauged Double Field Theory (GDFT).
The duality twisted dimensional reduction of the DFT action of the
RR sector with the anzats (\ref{anzats}) has been recently studied
by \cite{Aybike}.  Below we present briefly the reduced actions,
gage transformation rules and the consistency conditions for the
reductions. For more details, see
\cite{Geissbuhler,Aldazabal,Grana,Aybike}.

The reduced theory is determined by the so called fluxes $f_{ABC},
\ \eta_A$, which are defined as below: \be \label{structure}
f_{ABC} = 3 \Omega_{[ABC]}, \ \ \ \eta_A =
\partial_M  (U^{-1})^M_{\ A} - 2(U^{-1})^M_{\ A} \partial_M \rho, \ee  \be \label{omega}
\Omega_{ABC} = -(U^{-1})^M_{\ A} \del_M (U^{-1})^N_{\ B} U^D_{\
N}\eta_{CD}.  \ee Here $U = (\rho(S))^{-1} = \rho(S^{-1})$, where
$\rho$ is the double covering homomorphism. Note that
$\Omega_{ABC}$ are antisymmetric in the last two indices:
$\Omega_{ABC} = - \Omega_{ACB}$. We also make the following
definition \be \label{fa} f_A = -\partial_M (U^{-1})^M_{\ A}  =
\Omega^C_{\ AC}. \ee

The conditions for the consistency of the reduction of the NS-NS
sector can be listed as below\footnote{Despite the condition
(\ref{orthogonality}), we will be including the $\eta^A$ terms in
the following formulas for completeness, as it might be possible
to relax this condition by considering warped compactifications
\cite{Geissbuhler,Grana}. For non-vanishing $\eta^A$ one also
needs to impose the  conditions $\eta^A f_{ABC} = 0$ and $ \eta^A
\del_A g(X) = 0$ for the consistency of both the NS-NS and the RR
sectors.} \cite{Geissbuhler,Aldazabal,Grana}: \be
\label{onemlikosul} f^A_{\ BC} \del_A g(X) = 0,  \ee \bea
\label{jacobi} f_{E[AB} f_{C]D}^{\ \ \ E} &=& 0,\\
\label{orthogonality} \eta^A  &=& 0. \eea In addition to the
above, the weak and the strong constraint has to be imposed on the
external space so that \be \label{external}
\partial_A \partial^A V(X) = 0, \ \
\partial_A V(X) \partial^A W(X) = 0 \ee for any fields or gauge
parameters $V, W$ that has dependence on the coordinates of the
external space only. The last but not the least,   all fluxes must
be constant for the consistency of the reduced theory. This then
ensures that the $Y$ dependence is completely integrated out in
the reduced theory. Surprisingly, it is not necessary to impose
the strong and the weak constraints in the internal space, that
is, one does not need to impose \be \label{internalstrong}
\partial^P U^A_{\ M}
\partial_P U^B_{\ N}, \ \ \ \del_P \partial^P U^A_{\ M}. \ee Therefore, the duality twisted anzats
(\ref{anzats})-(\ref{anzats2}) allows for a relaxation of the
 constraints on the total space.

The requirement of consistency of the reduction of the DFT of the
RR sector brings in one more condition: the fluxes should also
satisfy \be \label{extraconstraint} f_{ABC} f^{ABC} = 0. \ee
Without this extra condition the reduced action cannot be
invariant under the reduced gauge transformation rules. It is
known that the constraints of GDFT are in one-to-one
correspondence with the constraints of half-maximal gauged
supergravity \cite{Grana}. We would like to  remark that  the
extra condition (\ref{extraconstraint}) means the the gauged
supergravity we have obtained  is a truncation of  maximal
supergravity \cite{Aldazabal:2011yz}.

Once the conditions discussed above are imposed, one finds that
the theory that results from the duality twisted reduction of the
DFT of Type II strings is a consistent theory with the following
Lagrangian for the RR sector (we discuss the NS-NS sector in the
Appendix. Full details can be found, for example, in
\cite{Grana}.) \be \label{redLag} L_{{\rm red}} = \frac{1}{4}
 \langle
 \slashed{\nabla}\chi(X) + \bar{\alpha}, \ C^{-1} \ciftS (\slashed{\nabla}\chi(X) + \bar{\alpha}) \rangle. \ee Here, the Dirac operator $
\slashed{\nabla}  $ and the constant spinor field $\bar{\alpha}$
are defined as \be \label{hatchi} \slashed{\nabla}   \equiv
\slashed{\del} + \frac{1}{6} f_{ABC} \psi^A \ \psi^B \ \psi^C +
\frac{1}{2} \eta_B \psi^B \ , \ee  \be \label{baralfa}
\bar{\alpha} \equiv (\frac{1}{6} f_{ABC} \psi^A \ \psi^B \ \psi^C
+ \frac{1}{2} \eta_B \psi^B) \alpha. \ee It can be shown that the
reduced Lagrangian can be rewritten as \cite{Aybike} \be
\label{reduced1} L_{{\rm red}} = \frac{1}{4} \langle  F(X),
 C^{-1}
S_g^{-1} F(X) \rangle + \frac{1}{2} \langle F(X),  C^{-1} S_g^{-1}
\bar{\chi}_B \rangle
 +  \frac{1}{4} \langle \bar{\chi}_B,   C^{-1}
S_g^{-1}
 \bar{\chi}_B \rangle, \ee where we have defined\footnote{Here, $S_b$ is the
spin group element whose spinorial action on polyforms corresponds
to $B$ field shifts, as we will discuss in more detail in section
(\ref{fluxes}).} $F(X) = S_b \slashed{\del}\chi(X) =  e^{-B}
\wedge \slashed{\del}\chi(X)$ and $\bar{\chi}_B = S_b (\bar{\chi}
+ \bar{\alpha}) = e^{-B} \wedge (\bar{\chi} + \bar{\alpha})$, with
\be \label{barchi} \bar{\chi}(X) = ( \frac{1}{6} f_{ABC} \psi^A \
\psi^B \ \psi^C + \frac{1}{2} \eta_B \psi^B) \chi(X) \ee and
$\bar{\alpha}$ is as in (\ref{baralfa}). If the external
coordinates $X$ do not include any dual coordinates, then the
above Lagrangian is equivalent to the following one \be
\label{reduced2} L_{{\rm red}} = \frac{1}{4} F(X) \wedge * F(X)
 + \frac{1}{2}  F(X) \wedge * \bar{\chi}_B
 +  \frac{1}{4}  \bar{\chi}_B \wedge *
 \bar{\chi}_B, \ee where * is the Hodge duality operator with
 respect to the metric $g(X)$.

On the other hand, the constraint reduces  to \be
\label{redselfduality} \slashed{\nabla} \chi(X) + \bar{\alpha} = -
C^{-1} \ciftS \ (\slashed{\nabla} \chi(X) + \bar{\alpha}). \ee

The deformed gauge transformation rules for the reduced theory  in
the RR sector are \bea \label{defgauge} \hat{\delta}_{\hat{\xi}}
\chi & \equiv &\delta_{\hat{\xi}} \chi + \frac{1}{4} f^A_{\ \ BC}
\ \hat{\xi}_A \ \Gamma^B \Gamma^C (\chi + \bar{\alpha})
+\frac{1}{2} \eta^A \
\hat{\xi}_A (\chi + \bar{\alpha})  \\
\delta_{\hat{\xi}} \chi & \equiv & \hat{\xi}^A \del_A \chi +
\frac{1}{2} \del_B \hat{\xi}_C \Gamma^B \Gamma^C (\chi + \bar{\alpha}) \\
\label{defgaugelambda} \hat{\delta}_{\hat\lambda} \chi  & = &
 \big(\slashed\del + \frac{1}{6}
f_{ABC} \psi^A \ \psi^B \ \psi^C + \frac{1}{2} \eta_B \psi^B \big)
\hl. \eea Here, we have defined $\xi^M(X, Y) = (U^{-1})^M_{\ A}
\hat{\xi}^A(X)$, $ \lambda(X, Y) = e^{-\rho(Y)} S(Y) \hl(X) $, $
\delta_\xi \chi =  S(Y) \big(\hat{\delta}_{\hat{\xi}} \chi), $ and
$ \delta_\xi \lambda = S(Y) \big(\hat{\delta}_{\hat{\lambda}}
\chi). $

It was shown in \cite{Aybike} that the reduced Lagrangian
(\ref{redLag}) is invariant under the deformed gauge
transformations with parameter $\hat{\lambda}$, only when the
Dirac operator $\slashed{\nabla}$ is nilpotent. It was also shown
that the  the Dirac operator is nilpotent if and only if the extra
condition (\ref{extraconstraint}) is satisfied. On the other hand,
the constraints (\ref{onemlikosul})-(\ref{external}) required for
the consistency of  the NS-NS sector are sufficient to ensure the
gauge invariance of (\ref{redLag}) under gauge transformations
with parameter $\hat{\xi}$ and they also suffice for the deformed
gauge transformations close to form an algebra with respect to the
deformed C-bracket \cite{Aybike}.

We  would like to note that in finding the reduced Lagrangian and
the gauge transformations, the following identity, which follows
from the fact that the Lie algebras of $SO(d,d)$ and $Spin(d,d)$
are isomorphic  plays a crucial role \cite{Aybike}: \be
\label{isom} \Gamma^A \ (U^{-1})^M_{ \ \ A} S^{-1}
\partial_M \ S = \frac{1}{12} f_{ABC} \Gamma^A \ \Gamma^B \ \Gamma^C
-\frac{1}{2} f_B \Gamma^B. \ee

%%%%%%%%%%%%%%%%%%%%%%%%%%%%%%%%%%%%%%%%%%%%%%%%%%%%%%%%%%%

\section{Double Field Theory of Massive Type IIA Theory}\label{sectionmassive}

The DFT action reduces to the standard Type II supergravity action
in the supergravity frame $\tilde{\partial}^i = 0$. In the RR
sector,  the action reduces to
 the action of massless Type II theory written in the democratic
formulation \cite{Fukuma,Bergshoeff}, as was shown explicitly in
\cite{dftRR}. Here, one should fix the chirality of the spinor
field $\chi$ at the outset. Positive chirality gives rise to Type
IIB theory, whereas negative chirality yields Type IIA theory.
Conversely, if one writes the democratic formulation of the RR
action in terms of the Mukai pairing, which is a $Spin(n,n)$
invariant bilinear form on the space of differential forms, then
it is easily seen that it extends naturally to the action
(\ref{DFTaction1}) of \cite{HohmKwak}, as was discussed in
\cite{Aybike}.

The Lagrangians
 for the
democratic formulation of Type II theories
 can be
written as \cite{Aybike}:\be \label{mukaiaction1} L = \frac{1}{4}
\langle F^{\pm} , C^{-1} S_g^{-1} F^{\pm} \rangle. \ee Here, $F^+$
and $F^-$ are polyforms consisting of even and odd degree forms,
respectively. The choice with $F^+$ gives the Lagrangian for the
Type IIA theory, whereas  the choice with $F^-$ gives the
Lagrangian for the Type IIB theory. $C$ is the charge conjugation
matrix and $S_g^{-1}$ is the spin representative  of the inverse
metric $g^{-1}$ \cite{Aybike}. The spinorial action of $C^{-1}
S_g^{-1}$ on a spinor field $\varphi$ amounts to taking the Hodge
dual of it with respect to the metric $g$, when $\varphi$ is
regarded as a differential form\cite{dftRR,Aybike}. Now, let
$\chi$ be the chiral spinor field, which encodes the (modified) RR
gauge potentials and their Hodge duals. It was shown in
\cite{Aybike} that (\ref{mukaiaction1}) can be rewritten as \be
\label{mukaiaction} L = \frac{1}{4} \langle  \ \psi^i
\partial_i \chi, \ C^{-1} \ciftS \ \psi^i
\partial_i \chi \rangle. \ee The crucial point to note here is that
the spinor field $\chi$, as a differential form, is related to
$F^{\pm} $ that appears in the democratic action in the following
way: $ F^{\pm} = S_b \psi^i \del_i \chi$, where $F^+$ corresponds
to the negative chirality and $F^-$ corresponds to the positive
chirality of $\chi$. Then, we have (dropping the superscripts
$\pm$ for convenience): \be \label{mukaiaction2} L = \frac{1}{4}
\langle F, C^{-1} S_g^{-1} F \rangle = \frac{1}{4} \langle \ S_b
\psi^i
\partial_i \chi, \ C^{-1} S_g^{-1} S_b \ \psi^i
\partial_i \chi \rangle. \ee The right hand side of this equation is indeed equivalent to (\ref{mukaiaction}), as
one has
 $\ciftS = S_b^\dagger S_g^{-1}
S_b$, where $S_b^\dagger \equiv C S_b^{-1} C^{-1}$ \cite{dftRR,
Aybike}. Also, the Mukai pairing is invariant under the action of
$S_b$, that is, $\langle S_b \phi, S_b \phi \rangle = \langle
\phi, \phi \rangle$ for any spinor field $\phi$ \cite{Gualtieri}.

The action (\ref{mukaiaction}) has to be supplemented by the
following self-duality equation: \be \label{selfddemocratic}
\slashed{\partial} \chi = -C^{-1} \ciftS \slashed{\partial} \chi.
\ee This condition is equivalent to the duality condition
\cite{HohmKwak} \be \label{duality} F_{10-p} =
(-1)^{[\frac{p-1}{2}]}
* F_p. \ee Here $F_p$ are components of the polyform $F$ and
$[\frac{p-1}{2}]$ is the first integer greater than or equal to
$\frac{p-1}{2}$. When one imposes this relation to the field
equations derived from the action in the democratic formulation,
the field equations for the higher degree fields (which come from
the Hodge duals of the RR potentials) become equivalent to the
Bianchi identities of the RR potential fields. It is
straightforward to extend this Lagrangian to its DFT version. One
just allows the fields $\ciftS$ and $\chi$ to depend on the
winding coordinates as well as the standard ones so that $\chi =
\chi(x, \tilde{x})$ and $\ciftS = \ciftS(x, \tilde{x})$. Then the
Dirac operator
 $\psi^i
\partial_i$ should also be extended to $\slashed{\partial}$ in (\ref{diracop}) and the
Lagrangian becomes \be \label{mukaiactionfinal} L = \frac{1}{4}
\langle  \ \slashed{\partial} \chi, \ C^{-1} \ciftS \
\slashed{\partial} \chi \rangle. \ee The duality relation
(\ref{selfddemocratic}) also extends trivially. Both the action
and the self-duality relation are  manifestly $Spin(d,d)$
invariant, as the Mukai pairing is itself $Spin(d,d)$ invariant.
However, it is crucial to include the dual coordinates, as $\chi
\rightarrow S \chi$ implies $\slashed{\partial} \chi \rightarrow S
\slashed{\partial} \chi$ only when the dual coordinates are
introduced \cite{Aybike}\footnote{To demonstrate invariance under
$Spin(d,d)$, one should also use the fact that $ C^{-1} \ciftS S =
\pm S C^{-1} \ciftS$ for any $S \in Spin(d,d)$. For details, see
\cite{Aybike}.}.

The massive IIA action also includes a 0-form field strength $F_0$
\cite{Polchinski}.  In order to take this extra 0-form field
strength into account, one has to modify
\cite{Bergshoeff,HohmKwak} \be \label{massiveIIA} F
\longrightarrow F +  e^{-B} \wedge F_0. \ee The Hodge duality
relation between $F_0$ and $F_{10}$, which follows from
(\ref{duality}), along with the field equation $d * F_{10} = 0$
implies that $dF_0 = 0$, that is, $F_0 = m$ is constant. The
corresponding action is equivalent to the Romans massive IIA
supergravity action, again in the above sense: varying the
democratic action and imposing the duality constraint
(\ref{selfddemocratic}), one obtains the Bianchi identities and
the field equations for the p-form fields in Romans' theory. Note
that, regarding $F_0$ as a 0-form field strength is somewhat
artificial, as the gauge potential that gives rise to it would
have to be a (-1)-form field. In spite of this, such a (-1)-form
field was introduced formally in \cite{Lavrinenko}. More recently,
the existence of such a (-1)-form field was considered in
\cite{HohmKwak} within the context of DFT, where they interpreted
it as a 1-form field, which depends on the dual winding
coordinates of DFT. The anzats \cite{HohmKwak} chooses for the
spinor field $\chi$ in order to induce such a mass deformation is
as follows: \be \label{hohmkwakanzats} \chi (x ,\tilde{x}) =
\chi_0(x) + \chi_1(\tilde{x}), \ee where the only non-vanishing
component of the spinor field $\chi_1(\tilde{x})$ is its 1-form
part, and it depends on  one of the dual coordinates $\tilde{x}_1$
 linearly: $\chi_1(\tilde{x}) = m \tilde{x}_1$. The dependence
on the dual coordinate $\tilde{x}$, being only linear, drops out
when the anzats (\ref{hohmkwakanzats}) is plugged into
$\slashed{\del} \chi$ yielding $\slashed{\del} \chi(x,\tilde{x}) =
\psi^i \del_i \chi_0(x) + \psi_i \tilde{\partial}^i
\chi_1(\tilde{x}) = \psi^i \del_i \chi_0(x) + m$. Then, $S_b \
\slashed{\del} \chi(x,\tilde{x}) \rightarrow S_b \ (\psi^i \del_i
\chi_0(x) + m) = F  + e^{-B} \wedge F_0, $ where, as above, we
identify $F(x)$ with $S_b (\psi^i \del_i \chi_0(x))$ and $F_0$
with $m$. As a result, the linear dependence of the 1-form part of
the spinor field $\chi$ on the dual coordinate $\tilde{x}$ indeed
induces the mass deformation desired. Note that the anzats
(\ref{hohmkwakanzats}) of \cite{HohmKwak} violates the strong
constraint, although it respects the weak constraint. Also note
that the gauge transformation rule (\ref{gaugechi}) depends on
$\chi$ explicitly (unlike the DFT action, which only depends on
derivatives of $\chi$) and hence the linear dependence on
$\tilde{x}$ does not automatically drop from the  gauge algebra.
Therefore, the standard gauge transformation rules for $\chi$ that
we presented in the previous sections have to be reformulated, so
that the closure of the gauge algebra requires only the weak
constraint \cite{HohmKwak}.

In this paper, we will introduce a different anzats,  which will
also induce a deformation of the type above. More precisely, we
will consider whether an anzats of the form (\ref{anzatschi})  for
the spinor field $\chi$ might give rise to a massive deformation
of Type IIA theory. We will see that this is indeed possible for
certain choices of the twist matrix $U \in SO^+(d,d)$ and the
corresponding $S \in Spin^+(d,d)$. Interestingly, for one choice
of the twist matrix, our anzats violates both the strong and the
weak constraints.  The consistency of our anzats owes to the fact
that  the strong and the weak constraints are violated only in the
internal space. The possibility of relaxation of the constraints
of DFT in the NS-NS sector through duality twisted reduction
anzatse was first noticed in \cite{Aldazabal,Geissbuhler} and was
established rigorously in \cite{Grana}, as we discussed in section
(\ref{review}). The consistency of such reductions in the RR
sector was studied in \cite{Aybike}. We also would like to mention
that,  our anzats does not require reformulation of the gauge
transformation rules in the RR sector as in \cite{HohmKwak}, and
we have full control over the effects of our anzats in the NS-NS
sector.

As we saw in the previous section, the deformation induced by a
duality twisted anzats of the form (\ref{anzats},\ref{anzatschi})
 is determined completely by
 the fluxes $f_{ABC}$ and $\eta_A$. Therefore, we would like to compute these fluxes
 first, in order to see the form of the twist matrices that might
 give rise to massive deformations of Type IIA theory.

 %%%%%%%%%%%%%%%%%%%%%%%%%%%%%%%%%%%%%%%%%%%%%%%%%%%%%%%%%%%%%%%%55

\subsection{Fluxes}\label{fluxes}
Recall that the twist element $S(Y)$ must be in the identity
component of the spin group: $S(Y) \in Spin^+(d,d)$. Let us
briefly recall
 the classification of elements of $Spin^+(d,d)$ and their
spinorial action on spinor fields \cite{dftRR, Aybike}. The
identity component $Spin^+(d,d)$ of the spinor group $Spin(d,d)$
is obtained by exponentiating the generators of the Lie algebra
$so(d,d) \cong spin(d,d)$ in the spin representation. This gives
the  elements $S_B, S_{\beta}, S_A$ below, which act on a spinor
field $\alpha$ as follows: \bea \label{spinelemB} S_B: \ \ \alpha
&\longmapsto & e^{-B} \wedge \alpha =
(1 - B + \frac{1}{2} B \wedge B - \ldots ) \wedge \alpha, \\
\label{spinelembeta} S_{\beta}: \ \ \alpha &\longmapsto &
e^{\beta} \alpha = (1 + i_{\beta} + \frac{1}{2} i_{\beta}^2 +
\cdots ) \alpha,
\\
\label{spinelemA} S_A: \ \ \alpha &\longmapsto &
\frac{1}{\sqrt{det R}} (e^{A})^* \alpha. \eea   Here $B =
 \frac{1}{2} B_{kl} \psi^k
\wedge \psi^l$, and $ \beta = \frac{1}{2} \beta^{kl} \psi_k \wedge
\psi_l.$  Also, $i_{\beta} \alpha = \frac{1}{2} \beta^{ij}
i_{\psi_i} (i_{\psi_j} \alpha) $ and $R^* \alpha = R_j^{\ i}
\psi^j \wedge i_{\psi_i} \alpha $, with $ R = e^{A}$, which is the
usual action of $GL^+(d)$ on forms, where $GL^+(d)$ is the space
of (orientation preserving) linear transformations  of strictly
positive determinant. Here, we have identified the spinor fields
with (not necessarily homogenous) differential forms (or with
polyforms, as is commonly called in physics literature), as in
(\ref{spinorform}) \cite{Gualtieri,dftRR}.

Under the double covering homomorphism $\rho: Spin(d,d)
\longrightarrow SO(d,d)$, these elements project onto the
$SO^+(d,d)$ elements $h_B, h_{\beta}$ and $h_A$, respectively,
where \bea\label{oddelemB} h_B \ &=&
\begin{pmatrix} 1 & -B \\ 0 & 1
\end{pmatrix} \ , \quad B^T = - B \; , \\
\label{oddelembeta} h_{\beta} \
 &=& \begin{pmatrix} 1 &   0 \\ \beta &
1
\end{pmatrix}, \quad  \beta^T = - \beta\\
 \label{oddelemA} h_A \
 &=& \begin{pmatrix} e^A &   0 \\ 0 &
(e^{-A})^T
\end{pmatrix}  \ ,  \label{PinEl33} \eea The corresponding
elements $S$ and $h$ satisfy \be \label{mainwithgamma} \Gamma_N
h^N_{\ M} = S \Gamma_M S^{-1}, \ \ \ (h^{-1})^M_{\ N} \Gamma^N = S
\Gamma^M S^{-1}. \ee

When we choose the twist matrix $S$ as one of  $S =
S_B(y,\tilde{y}) $, \ $S = S_{A}(y,\tilde{y}) $ or $S =
S_{\beta}(y,\tilde{y}) $, we introduce the dependence on the
coordinates $y, \tilde{y}$ through the parameters $B =
 B(y,\tilde{y}), A = A(y,\tilde{y}), \ \beta = \beta(y,\tilde{y})$. In
order to see how the DFT action is deformed in the RR sector (as
well as the NS-NS sector), it is enough to calculate $f_{ABC}$ and
$\eta_A$ in (\ref{structure}) for the corresponding $SO^+(d,d)$
matrices $U^{-1}_B, U^{-1}_A, U^{-1}_{\beta}$, as we discussed in
section (\ref{review}).

Let us consider the general case \be \label{twistU} U^{-1} =
h_{\beta} \ h_A \
h_B = \left(\begin{array}{cc} e^A & e^A B\\
                              \beta e^A & \beta e^A B + (e^{-A})^T
                       \end{array}        \right). \ee

Computation of the corresponding fluxes $f_{abc}, f^a_{\ bc},
f^{ab}_{\ \ c}, f^{abc}$ \footnote{Note that, due to complete
antisymmetry of $f_{ABC}$ in its indices, these are the only
independent blocks out of the 8 possible combinations.} has been
carried out by various groups
\cite{Andriot,Andriotflux,Aldazabal,Geissbuhler2,Lust}. It is
common to refer to them as the H-flux, geometric flux, Q-flux and
R-flux \cite{Wecht} (in the order they have been listed above),
and we will follow the same terminology.

If we introduce our duality twist through the $Spin^+(d,d)$
element $S = S_{\beta} . S_A . S_{B}$, then the fluxes
corresponding to $U^{-1} = \rho(S_{\beta} . S_A . S_{B}) =
\rho(S_{\beta}) \rho(S_A) \rho(S_B) = h_{\beta} \ h_A \ h_B$ in
(\ref{twistU}) is computed from the below\footnote{Note that in
the aforementioned references, where they have also computed the
fluxes, the twist matrix is of the form \be  U^{-1} = h_{B} \ h_A
\
h_{\beta} = \left(\begin{array}{cc} B (e^{-A})^T \beta + e^A & B (e^{-A})^T \\
                             (e^{-A})^T \beta  &  (e^{-A})^T
                       \end{array}        \right). \ee This brings in some differences in the computation of fluxes.
                       However, our results agree with theirs when $\beta = 0$ and/or $B = 0$.}:
\begin{eqnarray}\label{omegaflux}
  \Omega_{abc}&=& R^i_{\ a} (
  \partial_i  +  R_j^{\ d}  R_i^{\ e }B_{de} \tilde{\partial}^j ) B_{bc} + \Gamma_{a[b}^{\ \ \ d}B_{c]d} + B_{ea} \Gamma^{ed}_{\ \ [b}B_{c]d} \\
  & & + \beta^{km} R_k^{\ d} R_m^{\ f} B_{ad} (R^i_{\ f} \del_i B_{bc} +  \Gamma_{f[b}^{\ \ \ e}
  B_{c]e}) + R^i_{\ a}R_k^{\ e} R_j^{\ f} B_{eb} B_{jf} \del_i
  \beta^{jk} \nonumber \\
  & & +  R^i_{\ a}R_k^{\ e} R_j^{\ f} B_{eb}
  B_{da}(\tilde{\del}^i + \beta^{li} \del_l )\beta^{jk} \;,
  \nonumber \\
  \Omega^a_{\ bc} &=& R_l^{\ a} R_m^{\ \ e} \beta^{ml}
  \Gamma_{e[b}^{\ \ \ d}B_{c]d} + \Gamma^{ae}_{\ \ [b}B_{c]e}
 +R_i^{\ a} ( \tilde{\del}^i+ \beta^{li} \del_l ) B_{bc} \nonumber \\
  & & -R_i^{\ a} R_k^{\ d} R_j^{\ e} B_{db} B_{ec} (
  \tilde{\del}^i+ \beta^{li} \del_l
  )\beta^{jk} \;,
    \\ \nonumber
     \Omega_{ab}^{\ \ c} &=& -\Gamma_{ab}^{\ \ \ c} + R_k^{\ b} R_m^{\ e} \ \beta^{mk} \Gamma_{eb}^{\ \ c}+ B_{ad} \Gamma^{dc}_{\ \
     b}-
     R^i_{\ a} R_k^{\  d} R_m^{\ c} B_{db} (\del_i + R_j^{\ e} R_i^{\ f} B_{ef} \tilde{\del}^j) \beta^{mk}\;, \\ \nonumber
  \Omega_{a \   c}^{\  b} &=&  -  \Omega_{ac}^{\ \ b} \\ \nonumber
  \Omega^{ab}_{\ \ c} &=& \Gamma^{ab}_{\ \ c} - R_l^{\ a} R_m^{\ \ d} B^{lm} \Gamma_{dc}^{\ \ \ b} + R_i^{\ a} R_k^{\ b} R_j^{\ d} B_{cd}( \tilde{\del}^i+
   \beta^{li} \del_l ) \beta^{jk}\;, \\ \nonumber
  \Omega_a^{\ bc} &=& R^i_{\ a}R_j^{\ b} R_k^{\ c} \del_i
  \beta^{jk} + R_i^{\ d} R_j^{\ c} R_k^{\ b} B_{ad} (\tilde{\del}^i
  + \beta^{li} \del_l) \beta^{jk}\;,
  \\ \nonumber
  \Omega^{a \ c}_{\ b} &=& -\Omega^{ac}_{\ \ b} \\ \nonumber
  \Omega^{abc} &=& R_i^{\ a} R_j^{\ b} R_k^{\ c} (\tilde\partial^i + \beta^{li} \partial_l)
  \beta^{jk} \nonumber
   \,,
\end{eqnarray} Here $R_i^{\ a} = (e^A)_i^{\ a}$ and $R^i_{\  a} =
((e^{-A})^T)^i_{\ a}$. Also we have defined \bea \label{geomflux}
\Gamma_{ab}^{\ \ c} &=& -R^i_{\ a} \partial_i R_j^{\ c} R^j_{\ b}
\\
\Gamma^{ab}_{\ \ c} & = & -R_i^{\ a}  \tilde\partial^i R_j^{\ b}
R^j_{\ c}. \eea

Now one can find the fluxes from (\ref{structure}). For example,
the Q-flux is computed as \be f_a^{\ bc} = \Omega_a^{\ bc} +
\Omega^{bc}_{\ \ a} + \Omega^{c \ b}_{\ a}, \ee and similarly for
the other fluxes.

%%%%%%%%%%%%%%%%%%%%%%%%%%%%%%%%%%%%%%%%%%%%%%%%%%%%%%%%%%%%%%%%%%%

\subsection{Determining the twist element   that gives rise to Massive
IIA}\label{mainsection}

The first and the obvious condition on the duality twisted anzats
that will yield a massive deformation of Type IIA theory is that
the external coordinates $X$ should not include any dual
coordinates. In order to determine the twist matrix $U(Y)$ and the
corresponding $S(Y)$, let us recall that  the whole deformation
induced by the duality twisted anzats is encoded in $\bar{\chi}$
and $\bar{\alpha}$ in (\ref{barchi}) and (\ref{baralfa}),
respectively. Therefore, we are interested in the twists for which
these spinor fields include a 0-form part. Recall that $\psi^i$
 act on a spinor field $\phi$
 (regarded as a differential form) by wedge product,
 whereas $\psi_i$ acts on it by contraction, see (\ref{spinorform}) and (\ref{gammaact}).
  Therefore, the only way to obtain a 0-form field through the action of $\frac{1}{6} f_{ABC} \psi^A \psi^B \psi^C$
   is \footnote{From now on we are taking $\eta^A = 0$
  as is necessary for the gauge invariance of the reduced action.}  to act
 by $\psi_i$ with lower indices only. Therefore, we are
 interested in the fluxes $f_{ABC}$, which have either 3 upper
 indices $f^{abc}$ (R-flux), or 2 upper indices and 1 lower index of the
 form $f_a^{\ ba}$ (trace of the Q-flux)\footnote{Note that this
 is different from $f^b$. We have $f_a^{\ ba} = \Omega_a^{\ ba} +
\Omega^{ba}_{\ \ a} + \Omega^{a \ b}_{\ a}$, whereas $f^b =
\Omega_A^{\ bA} = \Omega_a^{\ ba} + \Omega^{ab}_{\ \ a} =
\Omega_a^{\ ba} - \Omega^{a \ b}_{\ a}$.}.  As we are not
interested in the geometric flux and the H-flux, we can assume
that $\partial_m R_a^{\ i} = 0$, that is, the components of the
matrix $A$ do not depend on any of the standard coordinates $y$,
and we also take $B = 0$. This immediately ensures that the
geometric flux and the H-flux vanish, as can be seen easily from
the flux formulas presented above. The expressions for the
remaining flux components also simplify significantly. We have
\begin{eqnarray}\label{omegafluxsimple}
  \Omega^{ab}_{\ \ c} &=& -R_i^{\ a} R^j_{\ c} \tilde\partial^i  R_j^{\ b}
  \\\nonumber
  \Omega_a^{\ bc} &=& R^i_{\ a} R_{j}^{\ b} R_k^{\ c} \partial_i \beta^{jk}
  \\ \nonumber
  \Omega^{abc} &=& R_i^{\ a} R_j^{\ b} R_k^{\ c} (\tilde\partial^i + \beta^{li} \partial_l)
  \beta^{jk}.
\end{eqnarray}

%%%%%%%%%%%%%%%%%%%%%%%%%%%%%%%%%%%%%%%%%%%%%55

\subsubsection{Twists with $S(y,\tilde{y}) =
S_A(\tilde{y})$, $\beta = 0$}\label{Atwist}

Let us take the $Spin^+(10,10)$ twist matrix  $S(y,\tilde{y}) =
S_A(\tilde{y})$ as in (\ref{spinelemA}). We simplify matters by
taking $A$ to depend on one of the dual coordinates only, which we
call $\tilde{x}_1$\footnote{In other words, our only $Y$
coordinate is $\tilde{x}_1$. Recall that we have already excluded
all the dual coordinates from the external $X$ coordinates.}, and
we also take this dependence to be linear. Let $A(\tilde{x}_1) = A
\tilde{x}_1$, where $A$ is in $gl(10)$. Then $(U^{-1}) = h_A =
\rho(S_A)$, where $h_A$ is as in (\ref{oddelemA}) with
$e^{A(\tilde{x}_1)} = e^{A \tilde{x}_1}$. We calculate
$\Omega_{ABC}$ from (\ref{omegafluxsimple}) and find \be
\Omega^{ab}_{ \ \ c} = -(e^{A\tilde{x}_1})_1^{\ a} A_c^{\ b},
\qquad \Omega^{a \
 c}_{ \  b} =  +(e^{A\tilde{x}_1})_1^{\
a} A_b^{\ c}, \qquad \Omega_a^{\ bc} = 0. \ee This gives the
following fluxes
 \be
 f_a^{\ bc} = \Omega_a^{\ bc} + \Omega^{bc}_{\ \ a} + \Omega^{c \ \ b}_{\ a} =\Omega^{[bc]}_{\ \ \ a} = -R_i^{\ [b} R^j_{\ a} \tilde\partial^i
R_{j}^{\ c]}  = -(e^{A\tilde{x}_1})_1^{\ [b} A_a^{\ c]}.
 \ee Also note that $f_a = \Omega^c_{\ ac} + \Omega_{ca}^{\ \ c} = 0$
and $f^a = \Omega^{ca}_{\ \ c} + \Omega_c^{\ ac} = \Omega^{ca}_{\
\ c} = -(e^{A\tilde{x}_1} A)_1^{\ a}= -\tilde{\partial}^1 R_1^{\
a} = -\tilde{\partial}^1 (U^{-1})_1^{\ a} $, as it has to be.

Consider the following simple choice of the twist matrix $U$,
where $A$ induces an an SL(2,R) twist in the parabolic conjugacy
class along the 1st and 2nd directions so that the only non-zero
component of the matrix $A$ is $A_1^{\ 2} = -m$. Then we have
$R_i^{\ a} = (e^{A\tilde{x}_1})_i^{\ a} = \delta_i^{\ a} -m
\tilde{x}_1 \delta_i^{\ 1} \delta_2^{\ a} $. This then gives us
$f_1^{\ 12} = m, $ and $f^2 = m$. The non-vanishing flux $f^2$
contributes to $\eta^2$ as can be seen from (\ref{structure}) and
(\ref{fa}). This should be compensated by a non-trivial dilaton
anzats (\ref{anzatsdilaton}) with non-constant $\rho$. We will
discuss this in more detail shortly. For the time being, let us
assume that $\rho$ has been chosen so as to yield vanishing
$\eta^2$, as is required for gauge invariance. Then we
have\footnote{Had we had a trivial dilaton anzats
(\ref{anzatsdilaton}) with constant $\rho$ so that $\eta^2 = -f^2
= -m$, we would have had $\bar{\chi} =
 m \psi_1 \psi_2 \psi^1  \chi, \ \ \ {\rm and} \ \ \
\bar{\alpha} = m \psi_1 \psi_2 \psi^1 \alpha.$} \be \bar{\chi} =
 (m \psi_1 \psi_2 \psi^1 + \frac{1}{2} m \psi_2) \chi, \ \ \ {\rm and} \ \ \
\bar{\alpha} = (m \psi_1 \psi_2 \psi^1 + \frac{1}{2} m \psi_2)
\alpha. \ee  Let us remind the reader, once again, that the
spinorial action of the Clifford algebra elements $\psi_1$ and
$\psi_2$ on the spinor field $\chi$ (regarded as a differential
form) is by contraction along the $x^1$ and $x^2$ directions,
respectively and $\psi^1$ acts by wedge product. Therefore, this
deformation term can give rise to a constant 0-form in the
resulting theory, when we choose the constant spinor field
$\alpha$ appropriately. For example, if we choose the spinor field
$\alpha$ such that its 1-form part has non-zero component along
$x^2$ direction, with coefficient 1 and has no other p-form parts
that have a non-vanishing component along this direction, then we
have $\bar{\alpha} = -\frac{1}{2} m$, which is a constant 0-form.
It is also possible to induce a non-constant 0-form field through
the 1-form part of the spinor field $\chi$. On the other hand, for
certain cases, for instance when $\chi$ has no p-form components
with non-vanishing coefficients along the $x^2$ direction, we will
just have $\slashed{\nabla}\chi(X) = \slashed{\del} \chi(X)$, as
we will have $\bar{\chi} = 0$.

Now that we have seen that our choice of anzats can give rise to a
constant (and also non-constant) 0-form field, we now would like
to discuss the non-trivial reduction anzats for the dilaton field.
Such non-trivial anzats is needed in order to make sure that
$\eta^2 = 0$ in the presence of non-zero $f^2$, as was discussed
in the previous paragraph. Looking at the definition of $\eta_A$
in (\ref{structure}), we see that the equation $\eta_A = 0$ is
equivalent to \be \label{dilatoneqn}
\partial_M (U^{-1})^M_{\ A} - 2 (U^{-1})^M_{\ A} \partial_M
\lambda = 0. \ee Since we have $(U^{-1})_{ia} = (U^{-1})^{ia} = 0$
and $ \del_i (U^{-1})^i_{\ \ a} = 0 $ for our choice of the twist
matrix, the equation (\ref{dilatoneqn}) gives us the following two
sets of coupled equations \bea (U^{-1})^i_{\ a}
\partial_i \lambda
= 0 \\
 \tilde{\partial}^1(U^{-1})_1^{\ a}  - 2 (U^{-1})_i^{\ a}\tilde{\partial}^i \lambda
= 0. \eea It is easily checked that \be \lambda = - \frac{m
\tilde{x}_2}{2}
 \ee solves these equations. Therefore, the anzats for
the generalized dilaton is \be \label{lambda} d(X, Y) = d(X) -
\frac{m \tilde{x}_2}{2} . \ee This ensures that the consistency
condition (\ref{orthogonality}) is satisfied.  We find this
dilaton anzats particularly interesting, given that reductions
with non-trivial dilaton anzats have received some interest
recently, in an attempt to understand the recently discovered
generalized supergravity equations \cite{Tseytlin} within DFT
\cite{Yoshida2,Yoshida1,Samtleben}. On the other hand, the
condition of constancy of $f_{ABC}$ and the Jacobi identity
(\ref{jacobi}) are trivially satisfied. The only other condition
we have to check is whether (\ref{onemlikosul}) is satisfied or
not. In our case, imposing this condition immediately implies that
all fields in the GDFT, including the generalized metric and the
dilaton, must be independent of the coordinate $x^1$ and $x^2$ and
the theory is effectively a 8-dimensional theory.

The fact that our theory is effectively 8-dimensional is perhaps
not very surprising, when one recalls the work of
\cite{hullmassive}, which explains how massive Type II string
theory arises from
 M-theory. Here, one compactifies M-theory on a torus bundle B(A,R) over a
circle $S^1$  of radius $R$, also taking the zero volume limit $A
\rightarrow 0$. The fibers of the bundle is a 2-torus $T^2$ and
its modulus depends on the coordinates of the circle $S^1$, where
the mass parameter $m$ results from this dependence and as such
becomes a measure of the non-triviality of the bundle. Then, when
$m=0$, this gives a compactification on a trivial torus bundle
over a circle, that is a 3-torus, in the limit in which the volume
of the 3-torus shrinks to zero. Therefore, the description of
massive string theory in \cite{hullmassive} can be understood as a
deformation of an effectively 8-dimensional theory.

Note that due to non-vanishing $f_1^{\ 12}$, the NS-NS
 sector is also deformed. We give more details about the NS-NS
 sector in the Appendix (\ref{Appendix}).

\subsubsection{Twists with $S(y,\tilde{y}) =
S_{\beta}(\tilde{y})$, $A = 0$} \label{betatwist}

Let us now consider the duality twisted reduction of the theory by
the $Spin^+(10,10)$ matrix $S = S_{\beta}(\tilde{y})$, where
$S_{\beta}$ is as in (2.38) with $\beta = \beta(\tilde{y})$. Then
the corresponding $SO^+(10,10)$ matrix is $U = h_{\beta}$, where
$h_{\beta}$ is as in (2.35). Repeating the calculation above we
find that the only non-zero components of $\Omega$ are: \be
\Omega^{1bc} = \tilde{\del}^1 \beta^{bc} \ee One immediately sees
that the  non-zero components of $\beta^{1j}$ will not bring any
non-zero contribution to $f^{1bc}$, so without loss of generality
we can take $\beta^{1j} = 0.$ Also, $f^{1bc}$ are constant only if
$\beta$ is linear in $\tilde{x}$. For simplicity, we take the only
non-zero components of $\beta$ to be $\beta^{23} = -\beta^{32} = m
\tilde{x}_1$. Then we have \be f^{123} = f^{231} = f^{312} = m.
\ee This then gives \be \bar{\chi} = m \psi_1 \psi_2 \psi_3 \chi \
\ \ {\rm and} \ \ \ \bar{\alpha} = m \psi_1 \psi_2 \psi_3 \alpha.
\ee This  leads to a constant 0-form (and hence a mass
deformation), if we choose $\alpha$ such that the integral of the
3-form part of it over the 3-cycle along the directions
$x^1x^2x^3$ is non-vanishing (and constant as $\alpha$ is a
constant spinor field to begin with). It is also possible to
introduce a non-constant 0-form field through contractions of the
3-form part of the spinor field $\chi$. On the other hand, we may
also have $\bar{\chi} = 0$ and hence  $\slashed{\nabla} \chi(X) =
\slashed{\del} \chi(X)$ when, for example, the 3-form part of
$\chi$ has no non-vanishing component along any of these 3
directions. Note that the resulting theory is effectively
7-dimensional, as the constraint $f^{ABC}
\partial_A g(X) = 0$ implies that no fields in the theory is allowed to depend on the
coordinates $x^1, x^2, x^3$. For details on the deformation of the
NS-NS sector, see Appendix (\ref{Appendix}).

%%%%%%%%%%%%%%%%%%%%%%%%%%%%%%%%%%%%%%%%%%%%%%%%%%%%%%%%%%%%%%%%

\subsubsection{Mixed twists $S(y,\tilde{y}) =
S_{\beta}(y) . S_{A}(\tilde{y})$}\label{newtwist}

In the previous subsection (\ref{Atwist}), we introduced a twist
which produces non-zero flux $f^{12}_{\ \ 1}$ along with
non-vanishing $f^2$. Since $f^2$ contributes to $\eta^2$, which is
required to be zero, we had to cancel $f^2$ via  dependence of the
dilaton field on one of the dual coordinates. In this section, we
will consider a new type of twist, which we will dub "the mixed
twist", as it will involve both the standard coordinates $x$ and
the dual coordinates $\tilde{x}$. This new twist will be of the
form \be \label{mixedtwist} S = S(y, \tilde{y}) = S_{\beta}(y) .
S_A(\tilde{y}), \ee where $S_A$ is as in subsection
(\ref{Atwist}). Therefore, we will have non-vanishing $f^{12}_{\ \
1}$ and $f^2$ flux, as before. This time, the non-zero $f^2$ flux
will be cancelled not by a non-trivial dilaton anzats, but by the
contribution of $S_{\beta}$ in (\ref{mixedtwist}). Choosing
$S_{\beta}(y) $ such that the only non-vanishing component of
$\beta$ in (\ref{twistU}) is  $\beta^{12} =  mx^1$ does indeed
yield the desired configuration. One can compute from
(\ref{omegafluxsimple}) that the only non-vanishing components of
$\Omega$ are \be \nonumber \Omega^{12}_{\ \ 1} = -\Omega^{1 \
2}_{\ 1} =m, \quad \Omega^{22}_{\ \ 1} = -\Omega^{2 \ 2}_{\ 1} =
-m^2 \tilde{x}^1, \quad \Omega_1^{\ 12} = - \Omega_1^{\ 21} = m,
\quad \Omega^{212} = - \Omega^{221} = m^2 \tilde{x}^1. \ee They
combine to give the flux combination exactly of the form we are
seeking for:  $f_{1}^{\ 12} = 2m$ and  $\eta^2 = f^2 = 0$. Then
the deformation in the RR sector is exactly as in subsection
(\ref{Atwist}) (except for the insignificant difference that here
in this section we have $2m$ rather than $m$). An appropriate
choice of $\alpha$ induces a 0-form and hence a mass deformation
and consistency requires that the fields in the resulting GDFT
depend only on the standard coordinates $\{x^3, \ldots, x^{10}\}$.

Note that the twists we have considered so far all involve 10 or
less coordinates on the total space: the 10 coordinates $(Y, X) =
(\{\tilde{x}_1, \tilde{x}_2\}, \{x^3, \ldots, x^{10}\})$ in Case
1; the 8 coordinates $(Y, X) = (\{\tilde{x}_1\}, \{x^4, \ldots,
x^{10}\})$ in Case 2 and the 10 coordinates $(Y, X) =
(\{\tilde{x}_1, x^1\}, \{x^3, \ldots, x^{10}\})$ for the last
case. It is also possible to come up with a configuration, which
depends on a total of more than 10 coordinates on the total space
and yet reproduces the same gauging. Consider the twist
(\ref{mixedtwist}) again, where $S_A$ is again as in section
(\ref{Atwist}), and $S_\beta$ is as in this section but this time
with $\beta^{12} = 2m x^1$. Now we have \be \nonumber
\Omega^{12}_{\ \ 1} = -\Omega^{1 \ 2}_{\ 1} =m, \quad
\Omega^{22}_{\ \ 1} = -\Omega^{2 \ 2}_{\ 1} = -m^2 \tilde{x}^1,
\quad \Omega_1^{\ 12} = - \Omega_1^{\ 21} = 2 m, \quad
\Omega^{212} = - \Omega^{221} = 4 m^2 \tilde{x}^1. \ee They
combine to give the flux combination $f_{1}^{\ 12} = 3 m$ and $f^2
= -m$. The non-vanishing $f^2$ should again be compensated by a
non-trivial dilaton anzats. For the case at hand, for which only
$(U^{-1})_{ia} = 0$ ( and still $ \del_i (U^{-1})^i_{\ \ a} = 0
$), the equation (\ref{dilatoneqn}) corresponding to $\eta^A = 0$
gives the following set of equations: \bea (U^{-1})^i_{\ a}
\partial_i \lambda
= 0 \\
\del_1 (U^{-1})^{1a} + \tilde{\partial}^1(U^{-1})_1^{\ a} -
2(U^{-1})^{ia}
\partial_i \lambda - 2 (U^{-1})_i^{\ a}\tilde{\partial}^i \lambda
= 0. \eea It is easily checked that \be \lambda =  \frac{m
\tilde{x}_2}{2}  \ee solves these equations. Therefore, the anzats
for the generalized dilaton is \be \label{lambda} d(X, Y) = d(X) +
\frac{m \tilde{x}_2}{2}. \ee Although this configuration generates
the same type of gauging, the total number of coordinates that the
fields on the total space can depend on are different. This time,
the number of allowed coordinates is 11: $(Y,X) = (\{\tilde{x}_1,
\tilde{x}_2, x^1\}, \{x^3, \ldots, x^{10}\})$.  The first 3
coordinates $\{\tilde{x}_1, \tilde{x}_2, x^1\}$ are the
non-dynamical internal $Y$ coordinates and the remaining 8
coordinates are the dynamical external $X$ coordinates.

We would like to remark that this type of degeneracy in the
duality twisted reductions of DFT was also noted in \cite{Roest}.
Indeed, it was observed in \cite{Roest} that two inequivalent
twists may generate the same gauging. Moreover, they also showed
that it was possible for  one of these twists to be geometrical,
in the sense that it respects the strong and the weak constraints
and yet the other might be non-geometrical, violating both of  the
constraints in the internal doubled space. The phenomena we
observe here is exactly the same. Although they generate the same
gauging, the twist in section (\ref{Atwist}) and the mixed twists
we have considered here have a remarkable difference. The twist in
(\ref{Atwist}) respects both the weak and the strong constraints
on the total space. However, both constraints are violated in the
doubled internal space for the mixed twists we have considered
here. Indeed, \be \del^P \del_P U^A_{\ M} =0 \ee is not satisfied
by one of the components of the twist matrix $U$ in both cases.
This is most easily seen, when one writes down the twist matrix
explicitly, as we do in Appendix (\ref{AppendixB}) (for the first
example of this section with trivial dilaton anzats). The examples
we have studied here is a nice demonstration of the fact that the
weak constraint is stronger than the consistency conditions of
duality twisted reductions of DFT, as was discussed in
\cite{Grana}.

%%%%%%%%%%%%%%%%%%%%%%%%%%%%%%%%%%%%%%%%%%%%%%%%%%%%%%%%%%

\subsection{The Reduced Action}
In the previous section, we have identified the twists which give
rise to a deformation of the form \be \label{newchi}
\slashed{\partial}\chi(X, Y) \rightarrow
S(\tilde{\slashed{\nabla}} \chi(X) + F_0), \ee  where the 0-form
$F_0$ comes from contractions of $\alpha$ and $\chi$.\footnote{We
define $\tilde{\slashed{\nabla}}$ such that
$\tilde{\slashed{\nabla}} \chi(X) = \slashed{\nabla} \chi(X) -
G_0$, where $G_0$ is the 0-form field that comes from the
contractions of $\chi$.} This type deformation can occur only in
Type IIA, where $\chi$ has negative chirality (and hence
corresponds to a polyform consisting of odd differential forms),
as such deformations arose from contractions of a 1-form or a
3-form, which only exists in Type IIA. Also note that $F_0$ need
not be a constant form, as it might include a part coming from the
contractions of the spinor field $\chi$. In what follows, we will
assume that this is not the case (which is trivially satisfied if,
for example, $\chi(X,Y)$ has no components along the direction in
$x^2$, as discussed in the previous sections). Hence, $F_0$ is a
constant form. Furthermore, this assumption implies that
$\tilde{\slashed{\nabla}}\chi = \slashed{\nabla} \chi =
\slashed{\del} \chi. $

Let  us plug in (\ref{newchi}) and (\ref{anzats}) into the DFT
Lagrangian for the RR sector (\ref{mukaiactionfinal}).  \bea L &=&
\frac{1}{4} \langle S(Y) \left(
 \slashed{\partial}\chi(X)
+ F_0 \right), \ C^{-1} (S^{-1})^\dagger(Y)\ciftS S^{-1} (Y) S(Y)\left(\slashed{\partial}\chi(X) + F_0 \right) \rangle \nonumber \\
& = & \frac{1}{4} \langle S(Y) \left(
 \slashed{\partial}\chi(X)
+ F_0 \right), \ C^{-1}(S^{-1})^\dagger(Y) S_b^\dagger S_g^{-1}
S_b \left(\slashed{\partial}\chi(X) + F_0 \right) \rangle
\nonumber \\ & = & \frac{1}{4} \langle S(Y) \left(
 \slashed{\partial}\chi(X)
+ F_0 \right), \ S(Y) S_b^{-1} C^{-1}  S_g^{-1} S_b
\left(\slashed{\partial}\chi(X) + F_0 \right) \rangle \nonumber \\
& = & \frac{1}{4} \langle  S_b \left(\slashed{\partial}\chi(X) +
F_0 \right), \ C^{-1} S_g^{-1} S_b
  \left(\slashed{\partial}\chi(X) +
F_0 \right) \rangle
\nonumber \\
& = & \frac{1}{4} (F + e^{-B} \wedge F_0) \wedge * (F + e^{-B}
\wedge F_0), \label{finalaction} \eea where $F = S_b
(\slashed{\partial}\chi(X))$. This is the Lagrangian of the
massive IIA theory \cite{Bergshoeff,HohmKwak}, as was discussed in
section (\ref{sectionmassive}).  Note that in the second line
above we used the definition $\ciftS = S_b^\dagger S_g^{-1} S_b$,
whereas in the third line we used $C^{-1} (S^{-1})^\dagger(Y) =
S(Y) C^{-1}$ and $ S_b^{-1} C^{-1} = C^{-1} S_b^\dagger $ (since
both $S(Y)$ and $S_b$ are elements of $Spin^+(d,d)$) and finally
in the fourth line we used the invariance property of the Mukai
pairing under $Spin^+(d,d)$ transformations.

The reduced self-duality condition (\ref{redselfduality}) gives
\be \slashed{\del} \chi(X) + F_0 = -C^{-1} \ciftS (\slashed{\del}
\chi(X) + F_0), \ee which, in terms of the p-form components
$F^m_{p}$ of the spinor field $\slashed{\del} \chi + F_0$, is
equivalent to \be \label{massiveduality} F^m_{8-p} =
(-1)^{[\frac{p-1}{2}]}
* F^m_p. \ee The field equations arising from (\ref{finalaction})
must be supplemented by (\ref{massiveduality}), where * is the
Hodge opearator with respect to the metric $g(X)$.  Note that the
top degree form in (\ref{massiveduality}) is an 8-form, as we
assume that the components of all forms along the directions $x^1,
x^2$ have been integrated out. Also note that, even for the case
of non-constant $F_0$, which could be generated by non-zero
contractions of $\chi$, the field equation $d
* F_8 = 0$ \footnote{Note that the field equation of the 7-form
potential must be  $d * F_8 = 0$, as the only possible coupling of
it can be with the B-field, which would give a 9-form that
vanishes in 8 dimensions. This is also what happens in  10
dimension, where the field equation of the 9-form potential gives
$d * F_{10} = 0$, whereas the other field equations will be of the
form $d * F_{2n} + dB \wedge * F_{2n +2} = 0$.}  would have forced
its dual $F_0$ to be constant, as discussed in section
(\ref{sectionmassive}).

\section{Conclusion and Outlook}\label{conclusion}

In this paper, we considered the possibility of obtaining massive
deformations of Type IIA theory through a duality twisted
reduction in the RR sector. This is motivated by a paper of Hohm
and Kwak \cite{HohmKwak}, where they obtain massive IIA within
DFT, via a linear dependence of the spinor field $\chi$ (which
encodes the p-form fields) on one of the dual coordinates. Here,
we allow all the fields in the theory to depend on (some of) the
dual coordinates through a duality twisted anzats. We show that a
0-form field strength (and hence a mass parameter) can be
generated for  certain choices of the twist element $S(\tilde{y},
y)$. The twist elements should be chosen such that they  deform
the differential form, which encodes the \emph{field strengths} of
the (modified) RR gauge potentials by  a 0-form. The gauge
potential of such a 0-form field could only be a (-1)-form. This
is in line with the interpretation of \cite{HohmKwak} that
(-1)-form are 1-forms depending on the dual coordinates.

One interesting aspect of our anzats is that for some choices of
the twist element $S(\tilde{y}, y)$, both the strong and the weak
constraints are violated explicitly in the internal doubled space.
Let us call such twists, following \cite{Roest},
\emph{non-geometric twists}. The natural question that arises is
 whether  the deformations/gaugings we obtain in
lower dimensions through such non-geometric twists can be obtained
from T-duals of a conventional compactification of supergravity.
In order to discuss this issue, it is useful to introduce  some
more terminology from \cite{Roest}, where gaugings obtained from
duality twisted reductions of DFT was classified. Two important
notions from \cite{Roest} are the \emph{twist orbit} and the
\emph{orbit of gaugings}. Twist orbit is defined to be the set of
twist matrices connected by T-duality transformations and
likewise, the orbit of gaugings is defined as the set of gaugings
that are related by duality transformations. The importance of
these definitions lie in the fact that two twists that lie in the
same twist orbit generate gaugings belonging to the same orbit of
gaugings and two theories that belong to the same orbit of
gaugings are physically equivalent\footnote{Note that the reverse
argument of the first part of this sentence is not necessarily
true. Two twists that lie in different twist orbits may generate
gaugings in the same orbit of gauging, as we will discuss below.
This possibility was also emphasized in \cite{Roest}.}. It is then
natural to call a orbit of gauging geometric, if it includes at
least one representative gauging that can be obtained through a
conventional compactification. Now the question we posed above can
be paraphrased using this terminology: "Do the theories that we
have obtained here belong to a geometric or a non-geometric
orbit?"

The condition (\ref{extraconstraint}) was identified in
\cite{Roest} as a criterion to label the orbits of gaugings as
geometric or non-geometric and it was shown that gaugings that do
not satisfy (\ref{extraconstraint}) are non-geometric and require
a truly doubled background in order to be lifted to  a
 compactification of DFT. On the other hand, a geometric twist,
 which  automatically satisfies (\ref{extraconstraint}), would
 always give rise to a theory that lies in a geometric orbit of gauging \cite{Roest}.
Indeed, if the strong and the weak constraints are not violated,
it is always possible to T-dualize to a frame in which the fields
and the gauge parameters of the theory have no dependence on the
dual coordinates \cite{HullZ4}. Accordingly, the twists we
considered in sections (\ref{Atwist}) and (\ref{betatwist}), being
geometric twists (as they do not violate the strong and weak
constraints of DFT even in the internal space), cannot  give rise
to a theory that lies in a non-geometric orbit of gauging, despite
the appearance  of the Q- and R- fluxes. What about the twists we
considered in section (\ref{newtwist})? Such twists belong to an
interesting class, as they violate the strong and the weak
constraints explicitly, and hence they are non-geometric, and yet
they satisfy the constraint (\ref{extraconstraint}). For the
example we encounter in section (\ref{newtwist}), we immediately
see that, the resulting theory belongs to a geometric orbit of
gauging, as the the deformation it gives rise to is exactly
 the same as the deformation in section (\ref{Atwist}). This is a
  a phenomenon, which has already been discussed and exemplified in
  \cite{Roest}. Indeed, it was observed in \cite{Roest} too that
  twists that belong to different twist orbits
may generate the same gauging, even when  one of these orbits is
geometric and  the other is non-geometric, violating both of  the
constraints in the internal doubled space. The phenomena we
observe here is exactly the same. Therefore, we conclude that all
the solutions we obtain here, which are massive deformations of
Type II supergravity theories with various type of fluxes belong
to a geometric orbit of gauging  and hence can be T-dualized to
conventional compactifications of Type II massive supergravity.

Another interesting aspect of our work is that, for some choices
of the twist $S(\tilde{y}, y)$, requirement of consistency forces
the dilaton field to pick up a linear dependence one one of the
dual coordinates, through an anzats of the form $\phi(X,Y) =
\phi(X) + \rho(\tilde{y})$. This makes us wonder about the
possible connections between the field equations following from
such reductions and the recently discovered generalized
supergravity equations \cite{Tseytlin}. The reason one might hope
for such a connection is the recent works
\cite{Yoshida2,Yoshida1,Samtleben}, where it was shown that
generalized supergravity equations can be obtained from DFT (in
\cite{Yoshida1}) and from exceptional field theory
(EFT)\footnote{Exceptional Field Theory is  a  U-duality covariant
extension of supergravity.} (in \cite{Samtleben}) through an
anzats by which the dilaton acquires a linear dependence on the
dual coordinates. On the other hand, it is interesting to note
that the dilaton $\phi$ in the ordinary Type IIA supergravity
action (in the string frame) has  a shift symmetry, which also
allows the following Scherk-Schwarz type anzats for $\phi: \
\phi(x,y) = \phi(x) + \rho(y)$, where $y$ are the coordinates of
the internal manifold. In the papers \cite{Derendinger,CatalOzer},
such a reduction to 4 dimensions with liner dependence on $y$ was
considered. The resulting theory is a massive, gauged theory and
its Lagrangian can be put in the general form given by Sch\"{o}n
and Weidner in \cite{SchonWeidner}, where (part of) the $SL(2)$ of
the global $SL(2) \times SO(6,6)$ group was gauged. It would be
interesting to explore the type of gaugings that would arise in
four dimensions through a duality twisted reduction  with twists
involving $\rho = \rho(y, \tilde{y})$.

We would like to mention that massive deformations of Type IIA
theory has also been studied by various groups within the context
of EFT \cite{Ciceri, Cassani}. In \cite{Cassani}, massive Type IIA
theory arises as a purely geometric solution of a consistent
deformation of EFT, which is called XFT (referring to X-deformed
EFT). The deformation of EFT is based on a modification of the
generalised Lie derivative by non-derivative terms of the form \be
\tilde{\cL}_\Lambda = \cL_\Lambda + \Lambda^M X_M, \ee where $X_M$
take values in the Lie algebra of the U-duality group. Acting on a
field in a representation of the U-duality group, it takes the
form $(X_M)_N^{\  P} = X_{MN}^{\ \ \ P}$. The deformation is
consistent only if $X_{MN}^{\ \ \ P}$ satisfy a set of
constraints. Namely, one should have $$X_{MP}^{\ \ \ R} X_{NR}^{\
\ \ Q} - X_{NP}^{\ \ \ R} X_{MR}^{\ \ \ Q} + X_{MN}^{\ \ \ R}
X_{RP}^{\ \ \ Q} \ \ \ {\rm and} \ \ \ X_{MN}^{\ \ \ P} \del_P =
0. $$ It is interesting to note that the duality twisted anzats we
have studied in this paper also induce non-derivative deformations
of a similar type on the Lie derivative and the Dirac operator,
and the fluxes $f_{MN}^{\ \ \ P}$ which determine this deformation
should obey exactly the same type of constraints, as listed in
(\ref{onemlikosul}) and (\ref{jacobi}).

\section*{Acknowledgments}
This work is supported by the Turkish Council of Research and
Technology (T\"{U}B\.{I}TAK) through the ARDEB 1001 project with
grant number 114F321, in conjunction  with the COST action MP1405
QSPACE.

\appendix

\section{The NS-NS sector}\label{Appendix}
In this appendix, we review briefly the duality twisted reductions
of the DFT of the NS-NS sector of string theory. Our review
follows closely \cite{Grana}. Let us begin by presenting the
generalized Ricci scalar $\cR(\cH, d)$, that determines the action
(\ref{DFTaction1}). \bea \label{ricci} \cR(\cH, d) & = & 4
\cH^{MN}
\partial_M
\partial_N d - \del_M \del_N \cH^{MN} - 4 \cH^{MN} \del_M d \del_N d + 4 \del_M \cH^{MN} \del_N d \\
& & + \frac{1}{8} \cH^{MN} \del_M \cH^{KL} \del_N \cH_{KL} -
\frac{1}{2} \cH^{MN} \del_M \cH^{KL} \del_K \cH_{NL} \nonumber \\
& & + \frac{1}{2} \del_M \varepsilon^a_{\ P} \del^M
\varepsilon^b_{\ Q} S_{ab} \eta^{PQ} \nonumber \eea Here $\cH$ is
the generalized metric and  $\varepsilon^a_{\ P} $ is the
generalized vielbein with $\cH_{MN} = \varepsilon^a_{\ M} S_{ab}
\varepsilon^b_{\ N}$, where $S_{ab} = diag(-1,1,\cdots, 1; -1,1,
 \cdots,1)$ is the planar metric. The term in the last line is not
in the original generalized metric formulation of DFT and vanishes
when the strong constraint is imposed \cite{Grana}. When the
strong constraint is satisfied, the action (\ref{DFTaction}) is
invariant under the following gauge transformations, which forms a
gauge algebra that is closed with respect to the C-bracket.
\bea\label{manifestH}
  \delta_{\xi}{\cal H}_{MN} \ &=& \  \widehat{\cal L}_{\xi} {\cal H}_{MN} \\
  & \equiv & \ \xi^{P}\partial_{P}{\cal H}_{MN}
  +\big(\partial_{M}\xi^{P} -\partial^{P}\xi_{M}\big)\,{\cal H}_{PN}
  +
  \big(\partial_{N}\xi^{P} -\partial^{P}\xi_{N}\big)\,{\cal H}_{MP}\;, \nonumber \\
  \delta d ~\ &=& \ \xi^M \partial_M d - {1\over 2}  \partial_M \xi^M
  \, \nonumber
 \eea

After applying the reduction anzats (\ref{anzats2}),
(\ref{anzatsdilaton}) the generalized Ricci scalar is deformed as
$\cR \rightarrow \cR_{{\rm def}}$ with \be \label{deformedRicci}
\cR_{{\rm def}} = \cR + \cR_f, \ee \bea \cR_f &=& -\frac{1}{2}
f^A_{\ BC} \cH_{BD} \cH^{CE} \del_D \cH_{AE} -\frac{1}{12} f^A_{\
BC} f^D_{\ EF} \cH_{AD} \cH^{BE} \cH^{CF}
\nonumber \\
& &  -\frac{1}{4} f^A_{\ BC} f^B_{\ AD} \cH^{CD} - 2 \eta_A \del_B
\cH^{AB} + 4 \eta_A \cH^{AB} \del_B d - \eta_A \eta_B \cH^{AB},
\eea which determines an action closed under the following
deformed gauge transformation rules
\bea \hat{\delta}_{\hat{\xi}} \cH_{AB} & = & \delta_{\hat{\xi}} \cH_{AB} -f_{ACD} \hat{\xi}^C \cH^D_{\ B} + f^D_{\ CB} \hat{\xi}^C \cH_{AD} \\
\hat{\delta}_{\hat{\xi}} d & = & \delta_{\hat{\xi}} d -
\frac{1}{2} \eta_A \hat{\xi}^A. \eea Here $\xi^M(X, Y) =
(U^{-1})^M_{\ A} \hat{\xi}^A(X)$, as in section (\ref{review}). As
emphasized in section (\ref{review}), the closure of gauge algebra
now requires the imposition of the strong and weak constraints
only in the external space. In the following subsections, we will
analyze the deformations induced in the NS-NS sector by the twists
in subsections (\ref{Atwist}) and (\ref{betatwist}).

\subsection{Twists with non-vanishing $ f^{12}_{\ \ 1}$ : }

Consider the twists in sections (\ref{Atwist}) and
(\ref{newtwist}). They all give rise to the same deformation with
the only non-vanishing flux component $f^{12}_{\ \ 1}$, with all
other flux components zero (except the ones, obviously, related to
$f^{12}_{\ \ 1} $ by permutations of the indices). Plugging
$f^{12}_{\ \ 1} = m $ in (\ref{deformedRicci}) one finds \be
\label{defRicci1} \cR_f  = -\frac{1}{2} m \cH_{[1}^{\ \ d}
\cH_{2]E}
\partial_d \cH_1^{\ E} - \frac{1}{6} m^2 \cH^{11} \left(\cH_{11}
\cH_{22} - (\cH_{12})^2\right) - \frac{1}{2} m^2 \cH_{22}. \ee One
can write this in terms of the metric and B-field elements by
using the following parametrization  of the generalized metric
 \bea \cH_{ab} & = & g_{ab} - B_{ac} g^{cd} B_{db} \\
\cH_a^{\ b} & = & B_{ac} g^{cb}, \qquad \cH^a_{\ b} = -g^{ac}
B_{cb} \nonumber \\
\cH^{ab} & = & g^{ab}. \nonumber \eea Recall that the fields in
the resulting theory cannot have dependence on the coordinates
$x^1, x^2$, so  $\partial_d$ is non-zero only if $d \neq 1,2$.

One can also compute the deformed gauge transformation rules and
finds the following \footnote{Note that we have omitted the hats
on the gauge parameters for simplicity, so that $\hat{\xi}^A =
(\tilde{\xi}_i, \xi^i)$.}
\begin{eqnarray*} \delta g_{11} & = & \cL_{\xi} g_{11} + 2m(g_{11} B_{12} \xi^1
- g_{12} \tilde{\xi}_1 +
g_{11} \tilde{\xi}_2), \\ \delta g_{12} & = & \cL_{\xi} g_{12} + 2m g_{12} B_{12} \xi^1 + m g_{12} \tilde{\xi}_2 - m g_{22} \tilde{\xi}_1, \\
\delta g_{22} & = & \cL_{\xi} g_{22} + 2m g_{22} B_{12} \xi^1, \\
\delta g_{ab} & = & \cL_{\xi} g_{ab} + m(g_{1(a} B_{b)2} - g_{2(a} B_{b)1})\xi^1, \qquad  a, b \neq 1,2 \\
\delta B_{12} & = & \cL_{\xi} B_{12} + \del_{[1} \tilde{\xi}_{2]} +  m B_{12} \tilde{\xi}_2 - m(g_{11} g_{22} - (g_{12})^2 - (B_{12})^2) \xi^1 \\
\delta B_{ab} & = & \cL_{\xi} B_{ab} + \del_{[a} \tilde{\xi}_{b]}-
m(g_{a1} g_{2b} - g_{a2} g_{1b} + B_{a1} B_{2b} - B_{a2} B_{1b} )
\xi^1, \qquad a, b \neq 1,2
\end{eqnarray*} where $\cL_{\xi} u_{ij} \equiv \xi^p \del_p u_{ij} + \del_i \xi^p u_{pj} + \del_j \xi^p u_{ip}.
$

\subsection{Twists with non-vanishing $ f^{123} :$}

Consider the twist in section (\ref{betatwist}). Recall that the
fluxes induced by this twist are $f^{123} = m$. Then the
deformation in the generalized Ricci scalar is \be \cR_f =
-\frac{1}{2} m \cH_{[1}^{\ \ d} \cH_2^{\ E} \del_d \cH_{3]E} -
\frac{1}{12} m^2 {{\rm det}} \cH_{ij}, \ee where ${{\rm det}}
\cH_{ij}$ is the determinant of the  $3 \times 3 $ matrix whose
components are $\cH_{ij}$ with $i,j = 1,2,3$. Note that the
direction $x^d$ appearing in $\del_d$ in the first term cannot
include the directions $x^i, i=1,2,3$, as consistency requires
that the fields of the reduced theory should not depend on these
coordinates. The deformed gauge transformation rules are
\begin{eqnarray*} \delta g_{ij} & = &  -m
g_{i[1}B_{2\underline{j}}\tilde{\xi}_{3]} - m
g_{j[1}B_{2\underline{i}}\tilde{\xi}_{3]} \\
\delta B_{ij} & = & -m(g_{i[1} g_{2 \underline{k}} + B_{i[1} B_{2
\underline{k}}) \tilde{\xi}_{3]}
\end{eqnarray*}
Here, the underlined indices are not to be antisymmetrized.

\section{The Mixed Twist Matrix}\label{AppendixB}
We present here the explicit form of the twist matrix of section
(\ref{newtwist}) (the first one with trivial dilaton anzats).
 \begin{eqnarray*} (U^{-1})^M_{\ A}
& = & \left( \begin{array}{cc} (U^{-1})_i^{\ a} & (U^{-1})_{ia} \\
             (U^{-1})^{ia} & (U^{-1})^i_{\ a} \end{array} \right) \\ &  & \\
             & = & \left( \begin{array}{cc} \left( \begin{array}{ccc} \left(\begin{array}{cc}
              1 & -m \tilde{x}_1 \\ 0 & 1
             \end{array}\right) &  \cdots & 0 \\
             \vdots & & \vdots \\
             0 & \cdots & 1 \end{array}\right) &  \left(\begin{array}{ccc}
             0  & \cdots & 0 \\
             \vdots & & \vdots \\
             0 &  \cdots &  0 \end{array}\right) \\
           \left(\begin{array}{ccc}
             \left(\begin{array}{cc} 0 & m x^1 \\ -m x^1 & m^2 x^1
             \tilde{x}_1
             \end{array}\right) &  \cdots & 0 \\
             \vdots & & \vdots \\
             0 & \cdots & 0 \end{array}\right) &
               \left(\begin{array}{ccc}
\left(\begin{array}{cc} 1 & 0 \\ m \tilde{x}_1 & 1
             \end{array}\right) &  \cdots & 0 \\
             \vdots & & \vdots \\
             0 & \cdots & 1 \end{array}\right) \end{array}\right).
             \end{eqnarray*}


\begin{thebibliography}{99}










%\cite{Hull:2009mi}
\bibitem{HullZ1}
  C.~Hull and B.~Zwiebach,
  ``Double Field Theory,''
  JHEP {\bf 0909} (2009) 099
  doi:10.1088/1126-6708/2009/09/099
  [arXiv:0904.4664 [hep-th]].
  %%CITATION = doi:10.1088/1126-6708/2009/09/099;%%
  %322 citations counted in INSPIRE as of 22 Mar 2017

  %\cite{Hull:2009zb}
\bibitem{HullZ2}
  C.~Hull and B.~Zwiebach,
  ``The Gauge algebra of double field theory and Courant brackets,''
  JHEP {\bf 0909} (2009) 090
  doi:10.1088/1126-6708/2009/09/090
  [arXiv:0908.1792 [hep-th]].
  %%CITATION = doi:10.1088/1126-6708/2009/09/090;%%
  %203 citations counted in INSPIRE as of 22 Mar 2017

  %\cite{Hohm:2010jy}
\bibitem{HullZ3}
  O.~Hohm, C.~Hull and B.~Zwiebach,
  ``Background independent action for double field theory,''
  JHEP {\bf 1007} (2010) 016
  doi:10.1007/JHEP07(2010)016
  [arXiv:1003.5027 [hep-th]].
  %%CITATION = doi:10.1007/JHEP07(2010)016;%%
  %248 citations counted in INSPIRE as of 22 Mar 2017


%\cite{Hohm:2010pp}
\bibitem{HullZ4}
  O.~Hohm, C.~Hull and B.~Zwiebach,
  ``Generalized metric formulation of double field theory,''
  JHEP {\bf 1008} (2010) 008
  doi:10.1007/JHEP08(2010)008
  [arXiv:1006.4823 [hep-th]].
  %%CITATION = doi:10.1007/JHEP08(2010)008;%%
  %255 citations counted in INSPIRE as of 22 Mar 2017




%\cite{Tseytlin:1990va}
\bibitem{Tseytlin1}
  A.~A.~Tseytlin,
``Duality symmetric closed string theory and interacting chiral
scalars,''
  Nucl.\ Phys.\ B {\bf 350} (1991) 395.
  doi:10.1016/0550-3213(91)90266-Z
  %%CITATION = doi:10.1016/0550-3213(91)90266-Z;%%
  %208 citations counted in INSPIRE as of 29 Mar 2017

%\cite{Tseytlin:1990nb}
\bibitem{Tseytlin2}
  A.~A.~Tseytlin,
  ``Duality Symmetric Formulation of String World Sheet Dynamics,''
  Phys.\ Lett.\ B {\bf 242} (1990) 163.
  doi:10.1016/0370-2693(90)91454-J
  %%CITATION = doi:10.1016/0370-2693(90)91454-J;%%
  %192 citations counted in INSPIRE as of 29 Mar 2017

    %\cite{Siegel:1993xq}
\bibitem{Siegel1}
  W.~Siegel,
  ``Two vierbein formalism for string inspired axionic gravity,''
  Phys.\ Rev.\ D {\bf 47} (1993) 5453
  doi:10.1103/PhysRevD.47.5453
  [hep-th/9302036].
  %%CITATION = doi:10.1103/PhysRevD.47.5453;%%
  %231 citations counted in INSPIRE as of 30 May 2017

  %\cite{Siegel:1993th}
\bibitem{Siegel2}
  W.~Siegel,
  ``Superspace duality in low-energy superstrings,''
  Phys.\ Rev.\ D {\bf 48} (1993) 2826
  doi:10.1103/PhysRevD.48.2826
  [hep-th/9305073].
  %%CITATION = doi:10.1103/PhysRevD.48.2826;%%
  %285 citations counted in INSPIRE as of 30 May 2017

%\cite{Siegel:1993bj}
\bibitem{Siegel3}
  W.~Siegel,
  ``Manifest duality in low-energy superstrings,''
  hep-th/9308133.
  %%CITATION = HEP-TH/9308133;%%
  %83 citations counted in INSPIRE as of 30 May 2017


  %\cite{Hull:2004in}
\bibitem{Hulleski1}
  C.~M.~Hull,
  ``A Geometry for non-geometric string backgrounds,''
  JHEP {\bf 0510} (2005) 065
  doi:10.1088/1126-6708/2005/10/065
  [hep-th/0406102].
  %%CITATION = doi:10.1088/1126-6708/2005/10/065;%%
  %368 citations counted in INSPIRE as of 29 Mar 2017

%\cite{Dabholkar:2005ve}
\bibitem{Hulleski2}
  A.~Dabholkar and C.~Hull,
  ``Generalised T-duality and non-geometric backgrounds,''
  JHEP {\bf 0605} (2006) 009
  doi:10.1088/1126-6708/2006/05/009
  [hep-th/0512005].
  %%CITATION = doi:10.1088/1126-6708/2006/05/009;%%
  %185 citations counted in INSPIRE as of 29 Mar 2017

  %\cite{Hull:2006va}
\bibitem{Hulleski3}
  C.~M.~Hull,
  ``Doubled Geometry and T-Folds,''
  JHEP {\bf 0707} (2007) 080
  doi:10.1088/1126-6708/2007/07/080
  [hep-th/0605149].
  %%CITATION = doi:10.1088/1126-6708/2007/07/080;%%
  %184 citations counted in INSPIRE as of 28 Mar 2017


%\cite{Hull:2007jy}
\bibitem{Hulleski4}
  C.~M.~Hull and R.~A.~Reid-Edwards,
  ``Gauge symmetry, T-duality and doubled geometry,''
  JHEP {\bf 0808} (2008) 043
  doi:10.1088/1126-6708/2008/08/043
  [arXiv:0711.4818 [hep-th]].
  %%CITATION = doi:10.1088/1126-6708/2008/08/043;%%
  %71 citations counted in INSPIRE as of 28 Mar 2017

    %\cite{Hitchin:2010qz}
\bibitem{Hitchin}
  N.~Hitchin,
  ``Lectures on generalized geometry,''
  arXiv:1008.0973 [math.DG]; \
  %%CITATION = ARXIV:1008.0973;%%
  %43 citations counted in INSPIRE as of 28 Mar 2017
%\cite{Hitchin:2004ut}
%\bibitem{Hitchin:2004ut}
  N.~Hitchin,
  ``Generalized Calabi-Yau manifolds,''
  Quart.\ J.\ Math.\  {\bf 54} (2003) 281
  doi:10.1093/qjmath/54.3.281
  [math/0209099 [math-dg]].
  %%CITATION = doi:10.1093/qjmath/54.3.281;%%
  %511 citations counted in INSPIRE as of 28 Mar 2017

    %\cite{Gualtieri:2003dx}
\bibitem{Gualtieri}
  M.~Gualtieri,
  ``Generalized complex geometry,''
  math/0401221 [math-dg].
  %%CITATION = MATH/0401221;%%
  %530 citations counted in INSPIRE as of 28 Mar 2017




%\cite{Hohm:2011dv}
\bibitem{dftRR}
  O.~Hohm, S.~K.~Kwak and B.~Zwiebach,
``Double Field Theory of Type II Strings,''
  JHEP {\bf 1109} (2011) 013
  doi:10.1007/JHEP09(2011)013
  [arXiv:1107.0008 [hep-th]].
  %%CITATION = doi:10.1007/JHEP09(2011)013;%%
  %95 citations counted in INSPIRE as of 28 Mar 2017

%\cite{Hohm:2011zr}
\bibitem{dftRRkisa}
  O.~Hohm, S.~K.~Kwak and B.~Zwiebach,
  ``Unification of Type II Strings and T-duality,''
  Phys.\ Rev.\ Lett.\  {\bf 107} (2011) 171603
  doi:10.1103/PhysRevLett.107.171603
  [arXiv:1106.5452 [hep-th]].
  %%CITATION = doi:10.1103/PhysRevLett.107.171603;%%
  %81 citations counted in INSPIRE as of 28 Mar 2017


\bibitem{park1}
  I.~Jeon, K.~Lee, J.~H.~Park and Y.~Suh,
  ``Stringy Unification of Type IIA and IIB Supergravities under N=2 D=10 Supersymmetric Double Field Theory,''
  Phys.\ Lett.\ B {\bf 723} (2013) 245
  doi:10.1016/j.physletb.2013.05.016
  [arXiv:1210.5078 [hep-th]].
  %%CITATION = doi:10.1016/j.physletb.2013.05.016;%%
  %53 citations counted in INSPIRE as of 28 Mar 2017



  %\cite{Jeon:2012kd}
\bibitem{park2}
  I.~Jeon, K.~Lee and J.~H.~Park,
  ``Ramond-Ramond Cohomology and O(D,D) T-duality,''
  JHEP {\bf 1209} (2012) 079
  doi:10.1007/JHEP09(2012)079
  [arXiv:1206.3478 [hep-th]].
  %%CITATION = doi:10.1007/JHEP09(2012)079;%%
  %49 citations counted in INSPIRE as of 06 Apr 2017


  %\cite{Hohm:2011cp}
\bibitem{HohmKwak}
  O.~Hohm and S.~K.~Kwak,
``Massive Type II in Double Field Theory,''
  JHEP {\bf 1111} (2011) 086
  doi:10.1007/JHEP11(2011)086
  [arXiv:1108.4937 [hep-th]].
  %%CITATION = doi:10.1007/JHEP11(2011)086;%%
  %64 citations counted in INSPIRE as of 28 Mar 2017


    %\cite{Romans:1985tz}
\bibitem{Romans}
  L.~J.~Romans,
``Massive N=2a Supergravity in Ten-Dimensions,''
  Phys.\ Lett.\  {\bf 169B} (1986) 374.
  doi:10.1016/0370-2693(86)90375-8
  %%CITATION = doi:10.1016/0370-2693(86)90375-8;%%
  %437 citations counted in INSPIRE as of 28 Mar 2017



\bibitem{SS} J.~Scherk and J.~H.~Schwarz, ``How to get masses from
extra dimensions", Nucl.\ Phys.\ B {\bf 153}, (1979) 61 . \
J.~Scherk and J.~H.~Schwarz, ``Spontaneous Breaking Of
Supersymmetry Through Dimensional Reduction,'' Phys.\ Lett.\ B
{\bf 82} (1979) 60.
%%CITATION = PHLTA,B82,60;%%

















  %\cite{Geissbuhler:2011mx}
\bibitem{Geissbuhler}
  D.~Geissbuhler,
  ``Double Field Theory and N=4 Gauged Supergravity,''
  JHEP {\bf 1111} (2011) 116
  doi:10.1007/JHEP11(2011)116
  [arXiv:1109.4280 [hep-th]].
  %%CITATION = doi:10.1007/JHEP11(2011)116;%%
  %110 citations counted in INSPIRE as of 28 Mar 2017

%\cite{Aldazabal:2011nj}
\bibitem{Aldazabal}
  G.~Aldazabal, W.~Baron, D.~Marques and C.~Nunez,
  ``The effective action of Double Field Theory,''
  JHEP {\bf 1111} (2011) 052
   Erratum: [JHEP {\bf 1111} (2011) 109]
  doi:10.1007/JHEP11(2011)052, 10.1007/JHEP11(2011)109
  [arXiv:1109.0290 [hep-th]].
  %%CITATION = doi:10.1007/JHEP11(2011)052, 10.1007/JHEP11(2011)109;%%
  %142 citations counted in INSPIRE as of 28 Mar 2017

%\cite{Grana:2012rr}
\bibitem{Grana}
  M.~Grana and D.~Marques,
  ``Gauged Double Field Theory,''
  JHEP {\bf 1204} (2012) 020
  doi:10.1007/JHEP04(2012)020
  [arXiv:1201.2924 [hep-th]].
  %%CITATION = doi:10.1007/JHEP04(2012)020;%%
  %115 citations counted in INSPIRE as of 28 Mar 2017


  %\cite{Geissbuhler:2013uka}
\bibitem{Geissbuhler2}
  D.~Geissbuhler, D.~Marques, C.~Nunez and V.~Penas,
  ``Exploring Double Field Theory,''
  JHEP {\bf 1306} (2013) 101
  doi:10.1007/JHEP06(2013)101
  [arXiv:1304.1472 [hep-th]].
  %%CITATION = doi:10.1007/JHEP06(2013)101;%%
  %85 citations counted in INSPIRE as of 28 Mar 2017



  %\cite{Cho:2015lha}
\bibitem{parktwist}
  W.~Cho, J.~J.~Fernández-Melgarejo, I.~Jeon and J.~H.~Park,
  ``Supersymmetric gauged double field theory: systematic derivation by virtue of twist,''
  JHEP {\bf 1508} (2015) 084
  doi:10.1007/JHEP08(2015)084
  [arXiv:1505.01301 [hep-th]].
  %%CITATION = doi:10.1007/JHEP08(2015)084;%%
  %10 citations counted in INSPIRE as of 28 Mar 2017


  %\cite{Berman:2013cli}
\bibitem{berman}
  D.~S.~Berman and K.~Lee,
  ``Supersymmetry for Gauged Double Field Theory and Generalised Scherk-Schwarz Reductions,''
  Nucl.\ Phys.\ B {\bf 881} (2014) 369
  doi:10.1016/j.nuclphysb.2014.02.015
  [arXiv:1305.2747 [hep-th]].
  %%CITATION = doi:10.1016/j.nuclphysb.2014.02.015;%%
  %40 citations counted in INSPIRE as of 06 Apr 2017











%\cite{Hassler:2014sba}
\bibitem{Lust}
  F.~Hassler and D.~L\"{u}st,
  ``Consistent Compactification of Double Field Theory on Non-geometric Flux Backgrounds,''
  JHEP {\bf 1405} (2014) 085
  doi:10.1007/JHEP05(2014)085
  [arXiv:1401.5068 [hep-th]].
  %%CITATION = doi:10.1007/JHEP05(2014)085;%%
  %29 citations counted in INSPIRE as of 28 Mar 2017

  %\cite{Blumenhagen:2014gva}
\bibitem{hassler1}
  R.~Blumenhagen, F.~Hassler and D.~Lüst,
  ``Double Field Theory on Group Manifolds,''
  JHEP {\bf 1502} (2015) 001
  doi:10.1007/JHEP02(2015)001
  [arXiv:1410.6374 [hep-th]].
  %%CITATION = doi:10.1007/JHEP02(2015)001;%%
  %40 citations counted in INSPIRE as of 16 Oct 2017

%\cite{Blumenhagen:2015zma}
\bibitem{hassler2}
  R.~Blumenhagen, P.~du Bosque, F.~Hassler and D.~Lust,
  ``Generalized Metric Formulation of Double Field Theory on Group Manifolds,''
  JHEP {\bf 1508} (2015) 056
  doi:10.1007/JHEP08(2015)056
  [arXiv:1502.02428 [hep-th]].
  %%CITATION = doi:10.1007/JHEP08(2015)056;%%
  %28 citations counted in INSPIRE as of 16 Oct 2017

  \bibitem{Aybike}
  A.~Catal-Ozer,
  ``Duality Twisted Reductions of Double Field Theory of Type II Strings,''
  JHEP {\bf 1709} (2017) 044
  doi:10.1007/JHEP09(2017)044
  [arXiv:1705.08181 [hep-th]].


  \bibitem{Aldazabal:2011yz}
  G.~Aldazabal, D.~Marques, C.~Nunez and J.~A.~Rosabal,
 ``On Type IIB moduli stabilization and N = 4, 8 supergravities,''
  Nucl.\ Phys.\ B {\bf 849} (2011) 80
  doi:10.1016/j.nuclphysb.2011.03.016
  [arXiv:1101.5954 [hep-th]]; \ \ %\cite{Dibitetto:2011eu}
%\bibitem{Dibitetto:2011eu}
  G.~Dibitetto, A.~Guarino and D.~Roest,
``How to halve maximal supergravity,''
  JHEP {\bf 1106} (2011) 030
  doi:10.1007/JHEP06(2011)030
  [arXiv:1104.3587 [hep-th]].



  %\cite{Wulff:2016tju}
\bibitem{Tseytlin}
  G.~Arutyunov, S.~Frolov, B.~Hoare, R.~Roiban and A.~A.~Tseytlin,
  ``Scale invariance of the $\eta$-deformed $AdS_5\times S^5$ superstring, T-duality and modified type II equations,''
  Nucl.\ Phys.\ B {\bf 903} (2016) 262
  doi:10.1016/j.nuclphysb.2015.12.012
  [arXiv:1511.05795 [hep-th]]; \


%\cite{Baguet:2016prz}
\bibitem{Samtleben}
  A.~Baguet, M.~Magro and H.~Samtleben,
  ``Generalized IIB supergravity from exceptional field theory,''
  JHEP {\bf 1703} (2017) 100
  doi:10.1007/JHEP03(2017)100
  [arXiv:1612.07210 [hep-th]].
  %%CITATION = doi:10.1007/JHEP03(2017)100;%%
  %4 citations counted in INSPIRE as of 28 Mar 2017


\bibitem{Yoshida2}
 Y.~Sakatani, S.~Uehara and K.~Yoshida,
  ``Generalized gravity from modified DFT,''
  arXiv:1611.05856 [hep-th].
  %%CITATION = ARXIV:1611.05856;%%
  %5 citations counted in INSPIRE as of 29 Mar 2017

  %\cite{Sakamoto:2017wor}
\bibitem{Yoshida1}
  J.~i.~Sakamoto, Y.~Sakatani and K.~Yoshida,
  ``Weyl invariance for generalized supergravity backgrounds from the doubled formalism,''
  arXiv:1703.09213 [hep-th]; \
  %%CITATION = ARXIV:1703.09213;%%
  %\cite{Sakatani:2016fvh}


\bibitem{Mukai}
S.~Mukai, ``Symplectic Structure of the Moduli Space of Sheaves on
an Abelian or K3 Surface,'' Invent. Math., 77:101–116, 1984.



  %\cite{Fukuma:1999jt}
\bibitem{Fukuma}
  M.~Fukuma, T.~Oota and H.~Tanaka,
  ``Comments on T dualities of Ramond-Ramond potentials on tori,''
  Prog.\ Theor.\ Phys.\  {\bf 103} (2000) 425
  doi:10.1143/PTP.103.425
  [hep-th/9907132].
  %%CITATION = doi:10.1143/PTP.103.425;%%
  %63 citations counted in INSPIRE as of 28 Mar 2017




%\cite{Bergshoeff:2001pv}
\bibitem{Bergshoeff}
  E.~Bergshoeff, R.~Kallosh, T.~Ortin, D.~Roest and A.~Van Proeyen,
  ``New formulations of D = 10 supersymmetry and D8 - O8 domain walls,''
  Class.\ Quant.\ Grav.\  {\bf 18} (2001) 3359
  doi:10.1088/0264-9381/18/17/303
  [hep-th/0103233].
  %%CITATION = doi:10.1088/0264-9381/18/17/303;%%
  %216 citations counted in INSPIRE as of 28 Mar 2017


        %\cite{Polchinski:1995mt}
\bibitem{Polchinski}
  J.~Polchinski,
 ``Dirichlet Branes and Ramond-Ramond charges,''
  Phys.\ Rev.\ Lett.\  {\bf 75} (1995) 4724
  doi:10.1103/PhysRevLett.75.4724
  [hep-th/9510017].
  %%CITATION = doi:10.1103/PhysRevLett.75.4724;%%
  %2409 citations counted in INSPIRE as of 06 Apr 2017


\bibitem{Lavrinenko}
  I.~V.~Lavrinenko, H.~Lu, C.~N.~Pope and K.~S.~Stelle,
  ``Superdualities, brane tensions and massive IIA / IIB duality,''
  Nucl.\ Phys.\ B {\bf 555} (1999) 201
  doi:10.1016/S0550-3213(99)00307-7
  [hep-th/9903057].
  %%CITATION = doi:10.1016/S0550-3213(99)00307-7;%%
  %36 citations counted in INSPIRE as of 22 Jun 2017



  %\cite{Andriot:2012wx}
\bibitem{Andriotflux}
  D.~Andriot, O.~Hohm, M.~Larfors, D.~Lust and P.~Patalong,
``A geometric action for non-geometric fluxes,''
  Phys.\ Rev.\ Lett.\  {\bf 108} (2012) 261602
  doi:10.1103/PhysRevLett.108.261602
  [arXiv:1202.3060 [hep-th]].
  %%CITATION = doi:10.1103/PhysRevLett.108.261602;%%
  %101 citations counted in INSPIRE as of 22 Jun 2017



%\cite{Andriot:2012an}
\bibitem{Andriot}
  D.~Andriot, O.~Hohm, M.~Larfors, D.~L\"{u}st and P.~Patalong,
  ``Non-Geometric Fluxes in Supergravity and Double Field Theory,''
  Fortsch.\ Phys.\  {\bf 60} (2012) 1150
  doi:10.1002/prop.201200085
  [arXiv:1204.1979 [hep-th]].
  %%CITATION = doi:10.1002/prop.201200085;%%
  %106 citations counted in INSPIRE as of 28 Mar 2017

%\cite{Shelton:2005cf}
\bibitem{Wecht}
  J.~Shelton, W.~Taylor and B.~Wecht,
  ``Nongeometric flux compactifications,''
  JHEP {\bf 0510} (2005) 085
  doi:10.1088/1126-6708/2005/10/085
  [hep-th/0508133].
  %%CITATION = doi:10.1088/1126-6708/2005/10/085;%%
  %307 citations counted in INSPIRE as of 28 Mar 2017

%\cite{Hull:1998vy}
\bibitem{hullmassive}
  C.~M.~Hull,
  ``Massive string theories from M theory and F theory,''
  JHEP {\bf 9811} (1998) 027
  doi:10.1088/1126-6708/1998/11/027
  [hep-th/9811021].
  %%CITATION = doi:10.1088/1126-6708/1998/11/027;%%
  %99 citations counted in INSPIRE as of 23 Jun 2017

 \bibitem{Roest}
  G.~Dibitetto, J.~J.~Fernandez-Melgarejo, D.~Marques and D.~Roest,
  ``Duality orbits of non-geometric fluxes,''
  Fortsch.\ Phys.\  {\bf 60} (2012) 1123
  doi:10.1002/prop.201200078
  [arXiv:1203.6562 [hep-th]].
  %%CITATION = doi:10.1002/prop.201200078;%%
  %82 citations counted in INSPIRE as of 28 Mar 2017



  %\cite{Derendinger:2007xp}
\bibitem{Derendinger}
  J.~P.~Derendinger, P.~M.~Petropoulos and N.~Prezas,
  ``Axionic symmetry gaugings in N=4 supergravities and their higher-dimensional origin,''
  Nucl.\ Phys.\ B {\bf 785} (2007) 115
  doi:10.1016/j.nuclphysb.2007.06.021
  [arXiv:0705.0008 [hep-th]].
  %%CITATION = doi:10.1016/j.nuclphysb.2007.06.021;%%
  %37 citations counted in INSPIRE as of 28 Mar 2017

  %\cite{CatalOzer:2011wu}
\bibitem{CatalOzer}
  A.~Catal-Ozer, C.~Deliduman and E.~Ulas Saka,
  ``A Massive S-duality in 4 dimensions,''
  JHEP {\bf 1112} (2011) 102
  doi:10.1007/JHEP12(2011)102
  [arXiv:1110.4974 [hep-th]].
  %%CITATION = doi:10.1007/JHEP12(2011)102;%%

%\cite{Schon:2006kz}
\bibitem{SchonWeidner}
  J.~Schon and M.~Weidner,
  ``Gauged N=4 supergravities,''
  JHEP {\bf 0605} (2006) 034
  doi:10.1088/1126-6708/2006/05/034
  [hep-th/0602024].
  %%CITATION = doi:10.1088/1126-6708/2006/05/034;%%
  %127 citations counted in INSPIRE as of 29 Mar 2017



  %\cite{Ciceri:2016dmd}
\bibitem{Ciceri}
  F.~Ciceri, A.~Guarino and G.~Inverso,
 ``The exceptional story of massive IIA supergravity,''
  JHEP {\bf 1608} (2016) 154
  doi:10.1007/JHEP08(2016)154
  [arXiv:1604.08602 [hep-th]].
  %%CITATION = doi:10.1007/JHEP08(2016)154;%%
  %23 citations counted in INSPIRE as of 26 Oct 2017

  %\cite{Cassani:2016ncu}
\bibitem{Cassani}
  D.~Cassani, O.~de Felice, M.~Petrini, C.~Strickland-Constable and D.~Waldram,
 ``Exceptional generalised geometry for massive IIA and consistent reductions,''
  JHEP {\bf 1608} (2016) 074
  doi:10.1007/JHEP08(2016)074
  [arXiv:1605.00563 [hep-th]].
  %%CITATION = doi:10.1007/JHEP08(2016)074;%%
  %21 citations counted in INSPIRE as of 26 Oct 2017



\end{thebibliography}
\end{document}